\documentclass{article}

\usepackage{arxiv}

\usepackage[utf8]{inputenc} 
\usepackage[T1]{fontenc}    
\usepackage{hyperref}       
\usepackage{url}            
\usepackage{booktabs}       
\usepackage{amsfonts}       
\usepackage{nicefrac}       
\usepackage{microtype}      
\usepackage{graphicx}
\usepackage[numbers,sort&compress]{natbib}
\usepackage{doi}

\usepackage{amsmath,amssymb}
\usepackage{subcaption}
\usepackage{multirow}
\usepackage{enumerate}
\usepackage{color}
\usepackage{xcolor}       
\usepackage{pifont}       
\usepackage{caption}

\usepackage{algorithm}
\usepackage{algpseudocode}
\floatname{algorithm}{Algorithm}

\usepackage{amsthm}

\theoremstyle{definition}

\newtheorem{lemma}{Lemma}
\newtheorem{assumption}{Assumption}

\newtheorem{theorem}{Theorem}

\graphicspath{ {./Figs/} }

\title{Federated Policy Gradient for Dynamic Consensus: Gradient Dominance and Topological Bounds}

\author{
\hspace{1mm}Maryam Babazadeh
\\
	\texttt{m.babazadeh@ieee.org} \\
}

\date{}

\renewcommand{\headeright}{}
\renewcommand{\undertitle}{}

\hypersetup{
pdftitle={A template for the arxiv style},
pdfsubject={q-bio.NC, q-bio.QM},
pdfauthor={David S.~Hippocampus, Elias D.~Striatum},
pdfkeywords={First keyword, Second keyword, More},
}

\begin{document}
\maketitle

\begin{abstract}
This paper presents a federated optimization framework for dynamic consensus in homogeneous multi-agent systems. The underlying problem is challenged by the communication graph's spectral properties, which induce spatial heterogeneity and complicate decentralized control even when the dynamics of the agents are completely known. The problem is addressed in this paper by bridging distributed control theory with federated policy gradients. Accordingly, a comprehensive optimization landscape for dynamic consensus is established through a modal decomposition formulation. Specifically, a parameterized suboptimality bound is characterized that explicitly decouples the true aggregate policy gradient from the network covariance mismatch. Building on this formulation, sufficient conditions based on the graph's Laplacian spectrum are derived to recover exact gradient dominance, which guarantees convergence to the global consensus policy and establishes a lower bound on the achievable dominance coefficient. Furthermore, a Federated model-free consensus algorithm is proposed that utilizes zeroth-order exploration alongside a bisection-based discount factor annealing mechanism, successfully eliminating the need for an initial stabilizing policy. This approach achieves linear global convergence for policy optimization subroutines while maintaining strictly bounded outer iterations until dynamic consensus is achieved. Through extensive simulation studies and comparative feasibility analyses across diverse network topologies, it is demonstrated that the proposed federated approach consistently outperforms existing model-based and data-driven separation-based techniques.
\end{abstract}


\section{Introduction}
Dynamic consensus is a fundamental requirement for multi-agent systems, enabling individual agents to synchronize their behavior entirely through local communication. However, this coordination is strictly governed by the network's structural properties. The connectivity of the communication graph dictates both whether consensus is mathematically possible and how quickly the system converges. While continuous-time consensus is often achieved with only a spanning tree and controllable agents \cite{dutta2022strict}, far more restrictive conditions are required for discrete-time consensus. Its feasibility is governed by bounded consensus regions tied directly to the graph’s spectral properties, rendering the design of stabilizing feedback and coupling gains notoriously difficult for large networks with wide Laplacian eigenvalue spreads.

Methodologically, direct model-free control is established as an alternative to the traditional two-stage paradigm of system identification followed by model-based control. While cascading error propagation is frequently encountered in indirect control design methods, this limitation is avoided by model-free mechanisms. By relying on raw input-output data, the need for traditional system identification is bypassed entirely \cite{de2019formulas, farjadnasab2022model}. Yet, steep structural trade-offs are associated with existing model-free dynamic consensus approaches. In early schemes, agents were strictly assumed to be single-input single-output (SISO) \cite{yang2020cooperative, cao2023distributed}. Although feedback gains are learned directly from data in \cite{zhang2023data}, the approach relies on an unrealistic zero control-weighting assumption, and the feasibility limits imposed by the graph's spectrum are completely ignored. Furthermore, even model-free reinforcement learning (RL) methods, such as the Q-learning algorithm proposed in \cite{feng2021q}, are constrained to single-input systems and are hindered by a major practical hurdle: an initial stabilizing policy, derived either heuristically or from expert knowledge, is required prior to the initiation of learning. While the need for an initial stabilizing policy is eliminated and feedback and coupling gains are jointly designed via data-driven subroutines in \cite{TCNS2026}, significant conservatism is imposed by the approach, and a mechanism to adapt the global policy in the presence of mismatched agent dynamics is lacking. Consequently, as modern networks grow in complexity and operate in data-rich but model-poor environments, a critical need has emerged for model-free control paradigms capable of operating across decentralized architectures without requiring the sharing of raw trajectory data among agents.

By capturing the essential trade-offs of optimal control,  Linear Quadratic Regulation (LQR) serves as a rigorous foundational benchmark for evaluating the stability and convergence of model-free and RL algorithms. The intersection of RL and control theory has seen profound advancements, particularly following the demonstration that the LQR cost landscape, despite being non-convex in the controller parameters, exhibits a gradient dominance property (also known as the Polyak-Łojasiewicz condition) \cite{hu2023toward}. It was established in  \cite{fazel2018global} that this property guarantees global convergence for direct policy gradient methods in single-agent LQR. Subsequent research expanded on this foundation, detailing the sample complexity of zeroth-order (ZO)  methods \cite{malik2019derivative, mohammadi2021convergence}. Recent work \cite{fujinami2025policy} proves that for linear quadratic control under domain randomization, the policy gradient globally converges to a finite-sample approximation of the LQR objective, subject to appropriate heterogeneity bounds. While these studies solidify the theoretical viability of model-free continuous control, their analyses are confined to isolated agents or centralized architectures, where the objective function is uniformly governed by a single linear time-invariant system.

Federated learning (FL) has revolutionized collaborative learning by allowing $N$ agents to learn without sharing raw, sensitive data. Supported by distributed optimization, modern FL actively mitigates non-IID client drift \cite{lu2024federated} and overcomes network bottlenecks via asynchronous updates, adaptive sparsification, and secure evolutionary methods \cite{baduwal2026federated}. However, while FL is well-established in supervised tasks and has recently advanced in finite-MDP RL \cite{woo2025blessing}, its deployment in continuous control remains largely unexplored. This gap is fundamentally mathematical: unlike standard discrete MDPs where irreducibility and aperiodicity ensure convergence, physical dynamical systems mandate provable closed-loop stability and bounded trajectories across continuous state-action spaces, structural constraints, and non-convex optimization landscapes. Despite these challenges, recent studies have begun unlocking FL's potential in optimal control. In model-free LQR tracking, federating homogeneous agents reduces per-agent sample complexity by exploiting independent gradient noise, which also permits client subsampling \cite{ren2020federated}. Uplink overhead is further cut from $\mathcal{O}(d)$ (where $d$ is the policy dimension) to dimension-independent $\mathcal{O}(1)$ via scalar projections of zeroth-order gradients \cite{rostami2026scalar}. Although early work assumed identical dynamics for the decoupled agents \cite{ren2020federated, rostami2026scalar}, policy-gradient schemes have recently accommodated similar, non-identical LTI agents \cite{toso2024asynchronous, wang2026model} in a decoupled setup.

To overcome the conservativeness of available separation-based techniques for achieving dynamic consensus, there is a critical need for a framework that inherently respects the network's topological constraints during the learning process. This paper addresses this gap by reformulating the dynamic consensus problem through the lens of federated policy optimization. The primary theoretical challenge resolved herein pertains to gradient dominance: while this property is well-understood in the single-agent limit, it was previously entirely unclear how the spectral spread of the communication graph interacts with the policy gradient in a LQR-based dynamic consensus problem, and whether a shared consensus controller could provably converge to the global optimum. Accordingly, the main contributions of this work are threefold:
\begin{itemize}
    \item[(i)] The dynamic consensus problem is addressed for the first time through the lens of federated optimization. In contrast to existing studies on federated learning in control, which exclusively treat independent agents \cite{ren2020federated, rostami2026scalar, toso2024asynchronous, wang2026model}, a federated framework is established herein specifically for a consensus problem. Within this framework, an exact bounding technique for the networked LQR cost difference is derived, and an analytical slack parameter, $\epsilon > 0$, is introduced that explicitly decouples the true policy gradient norm from the network covariance mismatch. 
     \item[(ii)] It is proven that the reformulated optimization problem satisfies an approximate gradient dominance condition. A precise topological threshold is established (governed by the network's spectral spread and the LQR weighting matrices) under which the spatial residual is strictly dominated by the policy gradient. This guarantees convergence to the global consensus optimum without spurious stationary points and establishes the conditions on coupling graphs required to recover gradient dominance.
      \item[(iii)] A new algorithm is proposed to navigate agents into the optimal stabilizing basin through federation, completely eliminating the restrictive requirement for an initial stabilizing policy. The algorithm features a bisection-based discount factor annealing mechanism \cite{perdomo2021stabilizing}. By adaptively inflating the discount factor $\gamma \to 1$ while managing local zeroth-order exploration, the algorithm systematically transitions from stable, heavily discounted learning to exact optimal consensus, yielding a federated policy gradient method with guaranteed linear convergence to the optimal solution.
\end{itemize}

\textit{Notation:} The sets of $n$-dimensional real vectors and $n \times m$ real matrices are denoted by $\mathbb{R}^n$ and $\mathbb{R}^{n \times m}$, respectively. We define $\mathbb{S}^n$, $\mathbb{S}_{+}^n$, and $\mathbb{S}_{++}^n$ as the spaces of $n \times n$ symmetric, symmetric positive semi-definite, and symmetric positive definite matrices. The $n \times n$ identity matrix is represented by $I_n$. For real symmetric matrices $X$ and $Y$, the inequality $X \succ Y$ (resp., $X \succeq Y$) indicates that $X - Y \in \mathbb{S}_{++}^n$ (resp., $\mathbb{S}_{+}^n$), while $X \prec Y$ and $X \preceq Y$ denote the strict and non-strict negative definiteness of the difference. The Kronecker product of two matrices is denoted by $\otimes$. Furthermore, $\text{Tr}(X)$ and $\Vert{}X\Vert{}_F$ denote the trace and the Frobenius norm of a matrix $X$. The spectral radius of a square matrix $M$ is indicated by $\rho(M)$, and its eigenvalues are denoted by $\lambda_i(M)$.

\section{Problem Formulation}
Consider a homogeneous linear discrete-time multi-agent system of $N$ nodes interacting over a directed graph $\mathcal{G} = (\mathcal{V}, \mathcal{E})$, where $\mathcal{V} = \{ v_1, v_2, \dots, v_N \}$ is the set of nodes, indexed by $\mathcal{N} = \{1, 2, \dots, N\}$, and $\mathcal{E} \subseteq \mathcal{V} \times \mathcal{V}$ is the set of directed edges. Each edge from node $j$ to node $i$ is represented as $(v_j, v_i)$, indicating that information flows from node $j$ to node $i$. The adjacency matrix associated with $\mathcal{G}$ is given by $A_{adj} = [a_{ij}] \in \mathbb{R}^{N \times N}$, where $a_{ij} = 1$ if $(v_j, v_i) \in \mathcal{E}$, and $a_{ij} = 0$ otherwise. If $(v_j, v_i) \in \mathcal{E}$, then node $j$ is a neighbor of node $i$, and the set of neighbors of node $i$ is denoted by $\mathcal{A}_i = \{ j \mid (v_j, v_i) \in \mathcal{E} \}$.

Define the in-degree matrix \( D = \operatorname{diag}(d_i) \in \mathbb{R}^{N \times N} \), where each \( d_i = \sum_{j \in \mathcal{A}_i} a_{ij} \), and the Laplacian matrix \({\mathcal{L}= D - A_{adj}} \). Consequently, \(\mathcal{L}\mathbf{1}_N = 0 \), where \( \mathbf{1}_N \) is the \( N \)-dimensional vector of ones. The graph \( \mathcal{G} \) is said to contain a spanning tree if there exists at least one node such that a directed path connects it to every other node in \( \mathcal{G} \). This condition implies that \( \lambda_1(\mathcal{L}) = 0 \) is a simple eigenvalue of the associated Laplacian matrix \( \mathcal{L} \), while the other eigenvalues \( \lambda_2(\mathcal{L}), \ldots, \lambda_N(\mathcal{L}) \) have positive real parts \cite{west.2001}.

\begin{assumption}\label{assumption:1}
    The directed graph \( \mathcal{G} \) contains a spanning tree. 
\end{assumption}

\begin{assumption}\label{assumption:2}
   The Laplacian matrix of the graph $\mathcal{G}$ has entirely real eigenvalues. This includes undirected graphs, weight-balanced directed graphs, directed acyclic graphs (DAGs), hierarchical graphs, and other topologies whose network structures yield real spectra.
\end{assumption}

Let the dynamics of each node be given by,
\begin{equation}\label{eq:xi}
   x_i(k + 1) = A x_i(k) + B u_i(k), \quad \forall i \in \mathcal{N},
\end{equation}
where \( A \in  \mathbb{R}^{n_x\times n_x} \) and \( B \in  \mathbb{R}^{n_x\times n_u} \). The multi-agent system \eqref{eq:xi} is given in compact form  by
\begin{equation}\label{eq:compact}
    x(k + 1) = (I_N \otimes A) x(k) + (I_N \otimes B) u(k),
\end{equation}
where \( x(k) = \begin{bmatrix} x_1^T(k) & x_2^T(k) & \cdots & x_N^T(k) \end{bmatrix}^T \in \mathbb{R}^{n_xN} \) and \( u(k) = \begin{bmatrix} u_1^T(k) & u_2^T(k) & \cdots & u_N^T(k) \end{bmatrix}^T \in \mathbb{R}^{n_u N} \) represent the global state vector and the global control input, respectively. The primary objective is to design linear distributed controllers that achieve state synchronization among all agents. Specifically, this requires ensuring that$$\lim_{k \to +\infty} \Vert{} x_i(k) - x_j(k) \Vert{} = 0, \quad \forall i, j \in \mathcal{N}, \; i \neq .$$ This phenomenon is known as dynamic consensus and describes an emergent behavior where individual agent states asymptotically converge over time to align with the collective dynamics of the groups.

Under Assumption \ref{assumption:1} and the full column rank condition of matrix $B$, available distributed control strategies adopt a control law of the form $u_i(k) = -c K \sum_{j \in \mathcal{A}_i} a_{ij} \left( x_i(k) - x_j(k) \right)$. This approach traditionally relies on a decoupled two-step design process: computing the local feedback gain $K$ via a Riccati-based framework, followed by tuning the scalar coupling gain $c$. Existing  methodologies within this paradigm, includes model-based designs \cite{yang2020cooperative}, Q-learning framework \cite{feng2021q}, and recent data-driven convex-concave decomposition aimed at reducing design conservatism \cite{TCNS2026}. However, in all these methods, finding a feasible coupling gain becomes increasingly difficult when the graph Laplacian possesses a wide eigenvalue spectrum. Furthermore, because agents independently design their local feedback and coupling gains, the network fails to tolerate even slight mismatches among subsystems. Additionally, standard decentralized learning methods treat agents in isolation, meaning data from one subsystem cannot be utilized to assist data-scarce or heterogeneous neighboring systems. To address these challenges, this paper investigates the consensus problem through the lens of federated policy optimization. Instead of enforcing a rigid structural decomposition between the feedback gain and the coupling gain, agents collaboratively learn a unified consensus policy. This cooperative distributed data-driven paradigm eliminates the feasibility bottlenecks of decoupled designs, does not require an initial stabilizing policy, and uses collective network data to optimize performance globally.

Accordingly, consider the linear static feedback control law,
\begin{equation}\label{eq:ui}
u_i(k) = -K \sum_{j \in \mathcal{A}_i} a_{ij} \left( x_i(k) - x_j(k) \right),
\end{equation}
where $K \in \mathbb{R}^{n_u \times n_x}$ denotes the feedback control gain matrix. Expressing the control inputs in a compact vector form yields $u(k) = - (\mathcal{L} \otimes K) x(k)$, which leads to the overall closed-loop system dynamics,
\begin{equation}\label{eq:closed}
x(k + 1) = \left[ I_N \otimes A - \mathcal{L} \otimes (B K) \right] x(k).
\end{equation}
Under Assumptions \ref{assumption:1} and \ref{assumption:2}, the communication graph contains a spanning tree, implying that $\lambda_1(\mathcal{L}) = 0$ is a simple eigenvalue of the Laplacian matrix $\mathcal{L}$, while all remaining eigenvalues $\lambda_2(\mathcal{L}), \dots, \lambda_N(\mathcal{L})$ are positive. In Section \ref{sec:3}, a policy gradient algorithm is developed to guarantee dynamic consensus, synthesizing the control policy of the form \eqref{eq:ui} through the lens of the consensus-disagreement mode decomposition.

\section{A Policy Gradient Algorithm via Disagreement Coordinates}\label{sec:3}
\subsection{Disagreement coordinates and performance objectives}
Consider the Jordan canonical decomposition of $\mathcal{L}$, given by $\mathcal{L} = V J V^{-1}$, where $V \in \mathbb{R}^{N \times N}$ is the matrix of generalized eigenvectors and $J \in \mathbb{R}^{N \times N}$ is the Jordan canonical form comprising blocks associated with each eigenvalue $\lambda_i(\mathcal{L})$.
To rigorously decouple the collective network behavior from individual consensus errors, consider the modal coordinate transformation $z(k) = (V^{-1} \otimes I_{n_x}) x(k)$, partitioning the transformed state vector as $z(k) = \begin{bmatrix} z_1^T(k) & z_2^T(k) & \cdots & z_N^T(k) \end{bmatrix}^T$ with $z_i(k) \in \mathbb{R}^{n_x}$. Accordingly, in the inverse mapping $x(k) = (V \otimes I_n)z(k) = \sum_{i=1}^N (v_i \otimes I_n)z_i(k)$, $v_i$ denotes the $i$-th column of $V$. For the zero eigenvalue $\lambda_1 = 0$, the corresponding right eigenvector is the vector of ones, $v_1 = \mathbf{1}_N$. Thus, the first term contributes $(\mathbf{1}_N \otimes I_n)z_1(k)$ uniformly across all agents, dictating the global average trajectory and establishing $z_1(k)$ as the consensus mode. Conversely, for $i = 2, \dots, N$, let $w_i^T$ denote the $i$-th row of $V^{-1}$ (the left eigenvector corresponding to $\lambda_i$). These left eigenvectors satisfy the orthogonality condition $w_i^T \mathbf{1}_N = 0$. Consequently, by introducing the network average state $\bar{x}(k) = \frac{1}{N} \sum_{j=1}^N x_j(k)$ and noting that $\sum_{j=1}^N w_{ij} \bar{x}(k) = 0$, each disagreement coordinate can be rewritten as,
\begin{equation*}
    z_i(k) = \sum_{j=1}^N w_{ij} \left( x_j(k) - \bar{x}(k) \right).
\end{equation*}
Because the transformation matrix $V^{-1}$ is full rank, the set of left eigenvectors $\{w_2, \dots, w_N\}$ spans the subspace strictly orthogonal to the vector of ones $\mathbf{1}_N$. Therefore, requiring $z_i(k) \to 0$ for all $i = 2, \dots, N$ forces every individual agent's relative deviation $x_j(k) - \bar{x}(k)$ to vanish asymptotically, which is equivalent to state consensus. Under this transformation, the aggregate system dynamics evolve as,
\begin{equation}z(k+1) = \left( I_N \otimes A - J \otimes BK \right) z(k).\end{equation}

According to the upper block-triangular structure of $J$, the system exhibits the expanded representation,
\begin{equation*}
    z(k+1) =  \begin{bmatrix} A - \lambda_1 BK & \ast & \cdots & \ast \\ 0 & A - \lambda_2 BK & \ddots & \vdots \\ \vdots & \ddots & \ddots & \ast \\ 0 & \cdots & 0 & A - \lambda_N BK \end{bmatrix} z(k),
\end{equation*}
where the off-diagonal blocks ($\ast$) arise from the superdiagonal entries of the Jordan blocks in $J$. Accordingly, define a set of virtual, decoupled modal states $\tilde{x}_i$ governed by the modal subsystems $\tilde{x}_i(k+1) = A \tilde{x}_i(k) + \lambda_i B{u}_i(k)$. The asymptotic suppression of disagreement modes $z_i$ via the  distributed policy \eqref{eq:ui} can be governed entirely by the asymptotic stability of these modal subsystems with decoupled policies $u_i=-K\tilde{x}_i$. Accordingly, the consensus objective can be framed as minimizing the average LQR cost across the modal network given by,
\begin{equation*}J_{con}(K) = \frac{1}{N-1} \sum_{i=2}^N J_i(K, \lambda_i),\end{equation*}
with each individual modal cost given by the undiscounted expectation, \begin{equation*}J_i(K, \lambda_i) \triangleq \mathbb{E} \left[ \sum_{t=0}^{\infty} \left( \tilde{x}_i(k)^T Q \tilde{x}_i(k) + u_{i}(k)^T R u_{i}(k) \right) \right].\end{equation*}

Dynamic consensus is guaranteed by minimizing $J_{con}(K)$, where disagreement states and control efforts are penalized by $Q \succ 0$ and $R \succ 0$, respectively. A discount factor $\gamma \in (0, 1]$ is introduced to shrink the spectral radii of the system matrices, where the discounted modal dynamics are defined as $A_i(\gamma) = \sqrt{\gamma} A$ and $B_i(\gamma) = \sqrt{\gamma} \lambda_i B$ \cite{lamperski2020computing, perdomo2021stabilizing}. Accordingly, even the trivial controller $K = 0$ yields a finite initial cost for sufficiently small initial discount factor  $\gamma_0$, providing a well-defined starting point for policy optimization. As the feedback gain is iteratively refined, $\gamma$ is progressively annealed toward $1$. Ultimately, the boundedness of the average consensus cost $J_{con}(K \mid \gamma)$ at $\gamma = 1$ formally certifies the achievement of asymptotic consensus.

Assuming the initial virtual modal states are drawn from a normal distribution with a covariance matrix $\Sigma = \mathbb{E}[\tilde{x}_i(0)\tilde{x}_i(0)^T]$, for any stabilizing feedback gain $K$, the infinite-horizon discounted cost for the $i$-th independent mode evaluates $J(K, \lambda_i \mid \gamma) = \text{Tr}(P_{K,\gamma}^i \Sigma)$, where $P_{K,\gamma}^i \in \mathbb{S}^{n_x}_{++}$ is the unique positive-definite solution of the discounted discrete-time Lyapunov equation,
\begin{equation} \label{eq:lyapunov}
    P_{K,\gamma}^i = Q + K^T R K + \gamma (A - \lambda_i BK)^T P_{K,\gamma}^i (A - \lambda_i BK).
\end{equation}
To formally evaluate the policy gradient for the network consensus cost across all active communication modes $i \in \{2, \dots, N\}$, we define the foundational linear quadratic operators for each decoupled mode. Let $X_{K}^i \in \mathbb{S}_{++}^{n_x}$ denote the stationary state covariance matrix for mode $i$ under a fixed stabilizing controller $K$. Throughout the remainder of this paper, we assume for analytical simplicity that $\Sigma = I_{n_x}$.
Accordingly, $X_{K}^i$ is the unique positive-definite solution to the discrete-time Lyapunov equation,
\begin{equation}
    X_{K}^i = I + A_{cl}^i(K) X_{K}^i A_{cl}^i(K)^T,
\end{equation}
where $A_{cl}^i(K) = A_i + B_i K$ represents the closed-loop dynamics for mode $i$. Moreover, the stationarity matrix $G_{K}^i \in \mathbb{R}^{n_u \times n_x}$ characterizes the local descent direction and optimality gap for mode $i$. It represents the deviation from the necessary stationarity condition, defined as,
\begin{equation}
G_{K}^i = (R + B_i^T P_{K}^i B_i)K - B_i^T P_{K}^i A_i.
\end{equation}
At the mode-specific optimal controller, $G_{K}^i$ evaluates exactly to zero. Building upon these modal definitions, the exact policy gradient for the aggregate multi-agent network is obtained by evaluating the expected cost across the graph's Laplacian spectrum. By mapping the gradient of the average cost to the sum of the local modal gradients, the gradient of the overall network consensus cost $J_{\text{con}}(K)$ with respect to the controller $K$ is given by,
\begin{equation}
    \nabla J_{\text{con}}(K) = \frac{1}{N-1} \sum_{i=2}^N 2 G_{K}^i X_{K}^i.
\end{equation}

\subsection{Theoretical Guarantees for Performance and Discount Inflation}
\label{subsec:policy_gradient_mechanism}

\begin{assumption}\label{assumption:3}
Let $\gamma \in (0, 1]$ be the discount factor and let $c > 2$ be a tunable design parameter. For a given stabilizing static feedback gain $K$, let $P_{K,\gamma}^i \in \mathbb{R}^{n_x \times n_x}$ denote the corresponding discrete-time Lyapunov matrix for mode $i \in \{2, \dots, N\}$. The spectral spread of the network communication graph is bounded relative to the worst-case modal Lyapunov matrix according to,
\begin{equation}\label{eq:assumption3}
    \lambda_N - \lambda_2 \leq \frac{\lambda_{\min}(Q)^{3/2} \sqrt{\lambda_{\min}(R)}}{c \, n_x \sqrt{\gamma} \|B\|_2 \max_{i \in \{2, \dots, N\}} \|P_{K,\gamma}^i\|_2^2}.
\end{equation}
\end{assumption}

To ensure that the average consensus cost reliably reflects the performance of every individual disagreement mode across the network, the following Theorem is presented.

\begin{theorem}\label{Jbound}
Under Assumptions \ref{assumption:1}--\ref{assumption:3}, for any stabilizing static feedback gain $K$ residing within a bounded sublevel set of the discounted consensus cost, the worst-case individual mode cost is strictly bounded relative to the aggregate network cost. Specifically, it universally holds that,
\begin{equation*}
    \max_{i \in \{2, \dots, N\}} J(K, \lambda_i \mid \gamma) \leq \frac{c}{c - 2} J_{con}(K \mid \gamma).
\end{equation*}
\end{theorem}

\textit{Proof:} Let $A_{cl}^i(K,\gamma) = \sqrt{\gamma}(A - \lambda_i B K)$ denote the discounted closed-loop state matrix, and define $P_{\max} = \max_{i \in \{2, \dots, N\}} \Vert P_{K,\gamma}^i \Vert_2$ as the worst-case modal Lyapunov matrix norm. The structure of \eqref{eq:lyapunov} immediately results in,
\begin{align}\label{eq:Abound}
 \nonumber   & \Vert K \Vert_2 \leq \sqrt{\frac{\Vert P_{K,\gamma}^i \Vert_2}{\lambda_{\min}(R)}} \leq \sqrt{\frac{P_{\max}}{\lambda_{\min}(R)}},\\
      &\Vert A_{cl}^i(K,\gamma) \Vert_2 \leq \sqrt{\frac{\Vert P_{K,\gamma}^i \Vert_2}{\lambda_{\min}(P_{K,\gamma}^i)}} \leq \sqrt{\frac{P_{\max}}{\lambda_{\min}(Q)}}.
\end{align}
Accordingly, for any pair of modes $i$ and $j$,
\begin{align*}
    & \Vert{}A_{cl}^i(K,\gamma) - A_{cl}^j(K,\gamma)\Vert{}_2\leq \sqrt{\gamma}(\lambda_N - \lambda_2) \Vert{}B\Vert{} \sqrt{\frac{P_{\max}}{\lambda_{\min}(R)}}.
\end{align*}
This perturbation drives the Lyapunov matrix difference $\Delta P^{ij} = P^i_{K,\gamma} - P^j_{K,\gamma}$ via the discrete-time Lyapunov relation $\Delta P^{ij} = A_{cl}^i(K,\gamma)^T \Delta P^{ij} A_{cl}^i(K,\gamma) + Y^{ij}$, where the perturbation driver matrix $Y_{ij}$ is bounded in 2-norm by,
\begin{equation}
    \Vert Y^{ij} \Vert_2 \leq \Vert P_{K,\gamma}^j \Vert_2 \Vert A_{cl}^i(K,\gamma) - A_{cl}^j(K,\gamma) \Vert_2 \left( \Vert A_{cl}^i(K,\gamma) \Vert_2 + \Vert A_{cl}^j(K,\gamma) \Vert_2 \right)\leq \frac{2 \sqrt{\gamma}(\lambda_N - \lambda_2) \Vert B \Vert_2}{\sqrt{\lambda_{\min}(Q)\lambda_{\min}(R)}} P_{\max}^2.
\end{equation}
Because $A_{cl}^i(K,\gamma)$ is strictly Schur stable, the solution for $\Delta P^{ij}$ can be expressed as an infinite series, which guarantees that the matrix deviation is bounded by $\Vert \Delta P^{ij} \Vert_2 \leq \frac{\Vert Y^{ij} \Vert_2}{\lambda_{\min}(Q)} \Vert P_{K,\gamma}^i \Vert_2$. Expanding this expression explicitly in terms of the system parameters, we obtain,
\begin{equation}\label{eq:deltaP}
    \Vert \Delta P^{ij} \Vert_2 \leq \left( \frac{2 \sqrt{\gamma} (\lambda_N - \lambda_2) \Vert B \Vert_2}{\lambda_{\min}(Q)^{3/2} \sqrt{\lambda_{\min}(R)}} \right) P_{\max}^3.
\end{equation}
Equation \eqref{eq:deltaP} establishes a worst-case bound on the deviation between any two modal Lyapunov matrices across the network. By enforcing Assumption \ref{assumption:3} into deviation bound \eqref{eq:deltaP} the perturbation is strictly constrained by,
\begin{equation}\label{eq:DeltaP2}
    \Vert \Delta P^{ij} \Vert_2 \leq \frac{2}{c \, n_x} P_{\max}.
\end{equation}
Let the index $m = \arg\max_{k \in \{2, \dots, N\}} J(K, \lambda_k \mid \gamma)$ designate the worst-case mode. Utilizing the standard trace property for $n_x$-dimensional matrices where $J(K, \lambda_k \mid \gamma) = \text{Tr}(P^k_{K,\gamma})$, the scalar cost deviation satisfies $J(K, \lambda_m \mid \gamma) - J(K, \lambda_j \mid \gamma) \leq n_x \Vert{}\Delta P^{mj}\Vert{}_2$. Substituting the matrix deviation bound \eqref{eq:DeltaP2}, yields,
\begin{align*}
    &  J(K, \lambda_m \mid \gamma) - J(K, \lambda_j \mid \gamma) \leq \frac{2}{c} P_{\max} \leq \frac{2}{c} J(K, \lambda_m \mid \gamma).
\end{align*}
Rearranging terms yields the point-wise bound between the worst-case mode and any mode $j$, $$J(K, \lambda_m \mid \gamma) \leq \left(\frac{c}{c - 2}\right) J(K, \lambda_j \mid \gamma).$$ Averaging this upper bound across all $N-1$ active modes establishes the final network performance guarantee,
\begin{align*}
    &\max_{i} J(K, \lambda_i \mid \gamma) \leq \frac{1}{N-1} \sum_{j=2}^N \left(\frac{c}{c - 2}\right) J(K, \lambda_j \mid \gamma) = \frac{c}{c - 2} J_{con}(K \mid \gamma),
\end{align*}
which completes the proof. \hfill $\blacksquare$

The introduction of the tunable parameter $c > 2$ establishes a  relationship between the topological flexibility of the communication network and the strictness of the performance guarantees. The parameter $c$ explicitly links the spectral spread bound in Assumption \ref{assumption:3} to the cost inflation factor in Theorem \ref{Jbound} and serves as an analytical tuning parameter. A smaller $c$ means that the framework can tolerate a larger spectral spread. This allows for highly sparse, or loosely connected graphs. Conversely, a larger $c$ strictly tightens the bound, requiring the graph's eigenvalues to be clustered closely together, which typically necessitates a highly dense or nearly complete communication topology.

To establish a principled mechanism for progressively annealing the discount factor without causing the performance costs to blow up, Theorem \ref{thm:gamma_inflation} bounds the permissible increase in $\gamma$ such that the modal costs remain strictly controlled.

\begin{theorem}\label{thm:gamma_inflation}
Let $P_{K,\gamma}^i$ denote the unique positive definite solution to the discounted discrete-time algebraic Lyapunov equation for mode $i \in \{2, \dots, N\}$ under a given stabilizing static feedback gain $K$ and an initial discount factor $\gamma \in (0, 1]$. If the discount factor is inflated to a new value $\bar{\gamma} > \gamma$ satisfying the upper bound,
\begin{equation}\label{eq:inflationbound}
    \dfrac{\bar{\gamma}}{\gamma} \leq \min_{i \in \{2, \dots, N\}} \left( 1 + \frac{\lambda_{\min}(Q)}{2 \left( \Vert P_{K,\gamma}^i \Vert_2 - \lambda_{\min}(Q) \right)} \right),
\end{equation}
then the infinite-horizon cost for every individual mode across the network will at most double, i.e., for all active disagreement modes $i \in \{2, \dots, N\}$,
\begin{equation}
    J(K, \lambda_i \mid \bar{\gamma}) \leq 2 J(K, \lambda_i \mid \gamma).
\end{equation}
\end{theorem}

\textit{Proof:} 
Let $i \in \{2, \dots, N\}$ be an arbitrary mode in the network. The upper-bound \eqref{eq:inflationbound} is equivalently represented as,
\begin{equation*}
    2 \left( \frac{\bar{\gamma}}{\gamma} - 1 \right) \Vert{}P^i_{K,\gamma}\Vert{}_2 - \left( 2 \frac{\bar{\gamma}}{\gamma} - 1 \right) \lambda_{\min}(Q) \leq 0,
\end{equation*}
which implies,
\begin{equation}\label{eq:nonpositive}
    2 \left( \frac{\bar{\gamma}}{\gamma} - 1 \right) P^i_{K,\gamma} - \left( 2 \frac{\bar{\gamma}}{\gamma} - 1 \right) (Q + K^T R K) \preceq 0.
\end{equation}
Let $\Phi_{\bar{\gamma}}: \mathbb{S}_+^{n_x} \to \mathbb{S}_+^{n_x}$ define the discrete-time Lyapunov operator for the inflated closed-loop dynamics on the cone of positive semi-definite matrices, $\Phi_{\bar{\gamma}}(X) \triangleq (Q + K^T R K) + A_{cl}^i(K,\bar{\gamma})^T X A_{cl}^i(K,\bar{\gamma})$.
The exact solution $P^i_{K,\bar{\gamma}}$ is the unique fixed point of this operator, satisfying $\Phi_{\bar{\gamma}}(P^i_{K,\bar{\gamma}}) = P^i_{K,\bar{\gamma}}$. To demonstrate that the cost at most doubles, we evaluate this operator at the candidate matrix $2 P^i_{K,\gamma}$ as follows,
\begin{align}\label{eq:phi1}
   \Phi_{\bar{\gamma}}(2 P^i_{K,\gamma}) = (Q + K^T R K)+ 2 \frac{\bar{\gamma}}{\gamma} A_{cl}^i(K,\gamma)^T P^i_{K,\gamma} A_{cl}^i(K,\gamma).
\end{align}
Substituting $\Phi_{{\gamma}}(P^i_{K,{\gamma}}) = P^i_{K,{\gamma}}$ into the \eqref{eq:phi1} gives,
\begin{equation*}
    \Phi_{\bar{\gamma}}(2 P^i_{K,\gamma}) = (Q + K^T R K) + 2 \frac{\bar{\gamma}}{\gamma} \left( P^i_{K,\gamma} - (Q + K^T R K) \right).
\end{equation*}
To establish an upper bound, the difference between the mapped candidate and the candidate itself is derived as,
\begin{align}\label{eq:phibar}
     & \Phi_{\bar{\gamma}}(2 P^i_{K,\gamma}) - 2 P^i_{K,\gamma} = 2 \left( \frac{\bar{\gamma}}{\gamma} - 1 \right) P^i_{K,\gamma} - \left( 2 \frac{\bar{\gamma}}{\gamma} - 1 \right) (Q + K^T R K) \preceq 0,
\end{align}
where the last inequality in \eqref{eq:phibar} is given based on \eqref{eq:nonpositive}.
By the Comparison Theorem for discrete-time algebraic Lyapunov equations \cite{lancaster1995algebraic}, the Lyapunov operator is monotonically order-preserving. Consequently, any matrix $X_0$ satisfying $\Phi_{\bar{\gamma}}(X_0) \preceq X_0$ acts as a strict upper bound for the true fixed-point solution. The exact solution under the inflated discount factor therefore satisfies:
$P^i_{K,\bar{\gamma}} \preceq 2 P^i_{K,\gamma}$.  Finally, because the objective cost is evaluated via the trace operator, $J(K, \lambda_i \mid \gamma) = \text{Tr}(P^i_{K,\gamma})$, and the trace is a monotonically increasing linear operator on the positive semi-definite cone, we conclude $J(K, \lambda_i \mid \bar{\gamma}) \leq 2 J(K, \lambda_i \mid \gamma)$.\hfill $\blacksquare$

Theorem \ref{thm:gamma_inflation} verifies that satisfying the inflation bound \eqref{eq:inflationbound} guarantees the cost for any mode will at most double. It is notable that the discount factor inflation bound established in Theorem \ref{thm:gamma_inflation} represents a significant structural relaxation compared to the existing bound in \cite{perdomo2021stabilizing} which restrict the allowable step size using bounds proportional to $\mathcal{O}(1/\Vert{}P_{K,\gamma}^i\Vert{}^4)$ or worse (Assuming $Q=I$). In contrast, our derived bound scales proportionally to $\mathcal{O}(1/\Vert{}P_{K,\gamma}^i\Vert{})$.  This first-order bound is highly advantageous. When employing dynamic discount factor inflation (e.g., in bisection searches or policy iteration schemes aiming for $\gamma \to 1$), our bound guarantees that the modal costs will at most double while mathematically permitting strictly larger discrete step sizes.

\begin{lemma}\label{lem:covariance_lipschitz}
Let $X_{K_\gamma^*}^i$ and $X_{K_\gamma}^i$ be the steady-state covariance matrices for the $i$-th disagreement mode under the optimal stabilizing controller $K_\gamma^*$ and a perturbed stabilizing controller $K$, respectively. Let $P_{K_\gamma}^i$ denote the corresponding cost matrix for the perturbed controller. Under the modal decomposition, the difference in the state covariances is strictly bounded by,
\begin{align}\label{eq:deltaX}
     & \|X_{K_\gamma^*}^i - X_{K_\gamma}^i\|_F  \leq \frac{2\sqrt{\gamma} \lambda_{i} \|B\|_2 \|X_{K_\gamma^*}^i\|_2 \|P_{K_\gamma}^i\|_2^{\frac{3}{2}}}{\lambda_{\min}(Q)^{\frac{3}{2}}} \|K_{\gamma} - K_\gamma^*\|_F.
\end{align}
\end{lemma}
The Proof of Lemma \ref{lem:covariance_lipschitz} is given in the Appendix A.

The next lemma shows that the suboptimality gap between any stabilizing static feedback gain and the optimal consensus policy is strictly bounded by the Frobenius norm of the true aggregate policy gradient and the modal covariance mismatch residual.
\begin{lemma}\label{lemma:subopt}
Let $K$ be any stabilizing shared static feedback gain for the networked system under a given discount factor $\gamma \in (0, 1]$, and let $K_\gamma^*$ denote the optimal consensus policy achieving the minimum discounted cost $J_{\text{con},\gamma}^*$. The suboptimality gap is strictly bounded by,
\begin{align*}
    & J_{\text{con}}(K \mid \gamma) - J_{\text{con},\gamma}^* \leq  \frac{1+\epsilon}{\lambda_{\min}(R)\min_i \lambda_{\min}(X_{K_\gamma^*}^i)} \left\| \nabla J_{\text{con}}(K \mid \gamma) \right\|_F^2 + \frac{1+\epsilon^{-1}}{\lambda_{\min}(R)\min_i \lambda_{\min}(X_{K_\gamma^*}^i)} \left\| \mathcal{X}(K, K_\gamma^*) \right\|_F^2.
\end{align*}
where $\epsilon >0$ is an arbitrary positive number and the residual is defined as $\mathcal{X}(K, K_\gamma^*) = \frac{1}{N-1} \sum_{i=2}^N G_K^i (X_{K_\gamma^*}^i - X_K^i)$, with $G_K^i$ representing the exact marginal policy gradient factor for mode $i$.
\end{lemma}
The Proof of Lemma \ref{lemma:subopt} is given in the Appendix B.
The term $\epsilon>0$ allows the analyst to flexibly adjust the distribution of the upper bound between the gradient sensitivity and the network consensus penalty $\mathcal{X}(K, K_\gamma^*)$.

\subsection{The proposed Algorithm}
In this section, a model-based distributed algorithm is proposed to achieve dynamic consensus without requiring an initially stabilizing policy, provided that specific topological conditions are satisfied. To establish formal theoretical guarantees, sufficient conditions on the spectral properties of the interconnection network are derived such that gradient dominance is recovered for the discounted modal consensus objective $J_{\text{con}}(K \mid \gamma)$. Subsequently, convergence criteria for the intermediate policy-gradient updates and an upper bound on the number of outer annealing iterations required for termination are established.
\begin{theorem}
\label{lem:grad_domination}
Suppose the controller $K$ satisfies the bounded suboptimality condition $J_{\text{con}}(K \mid \gamma) \leq M J_{\text{con},\gamma}^*$ for some $M>0$. If the network's spectral spread parameter $c$ satisfies the topological condition,
\begin{align}\label{eq:topology_condition}
    &   M^5 \left( \frac{\gamma\lambda_{N} \|B\|_2}{\lambda_{\min}(R) \lambda_{\min}(Q)^5} \right) \left( \frac{c}{c-2} \right)^7 (J_{\text{con},\gamma}^*)^7   \leq \frac{\left(\sqrt{\lambda_{\min}(R) \mu}-1\right)^2}{4\mu},
\end{align}
then  exact gradient domination holds with $\mu >\dfrac{1}{\lambda_{\min}(R)}$ for the current discount factor, i.e.,
\begin{equation}
    J_{\text{con}}(K \mid \gamma) - J_{\text{con},\gamma}^* \leq \mu \|\nabla J_{\text{con}}(K \mid \gamma)\|_F^2.
\end{equation}
\end{theorem}

\textit{Proof:}
The first step is to bound the term  $\|\mathcal{X}(K, K_\gamma^*)\|_F$ based on the condition \eqref{eq:topology_condition}. Note that, 
\begin{equation}\label{eq:Xkkstar}
    \|\mathcal{X}(K, K_\gamma^*)\|_F \leq \left( \max_i \|G_{K}^i\|_2 \right) \left( \max_i \|X_{K_\gamma^*}^i - X_{K}^i\|_F \right)
\end{equation}
The stationarity matrix is directly bounded by the modal cost matrix, $\|G_{K}^i\|_2 \leq \max_i \|P_{K}^i\|_2$. For the covariance shift, we use  Lemma \ref{lem:covariance_lipschitz}) to obtain,
\begin{align} \label{eq:cov_shift_sub}
    & \max_i \|X_{K_\gamma^*}^i - X_{K}^i\|_F \leq   \frac{2 \sqrt{\gamma}\lambda_{N} \|B\|_2 \max_i \|X_{K_\gamma^*}^i\|_2 \|P_{K}^i\|_2^{\frac{3}{2}}}{\lambda_{\min}(Q)^\frac{3}{2}} \|K - K_\gamma^*\|_F.
\end{align}

Moreover,
\begin{equation}
    \|K - K_\gamma^*\|_F \leq \frac{1}{\sqrt{\lambda_{\min}(R)\min_i\lambda_{\min}(X^i_{K_\gamma})}} \sqrt{J_{\text{con}}(K \mid \gamma) - J_{\text{con},\gamma}^*} \leq \frac{1}{\sqrt{\lambda_{\min}(R)}} \sqrt{J_{\text{con}}(K \mid \gamma) - J_{\text{con},\gamma}^*}.
\end{equation}
Next, we establish a strict upper bound for the optimal state covariance norm $\max_i \|X_{K_\gamma^*}^i\|_2$. Because the LQR cost traces are strictly bounded below by the state penalty, we have $J^i(K_\gamma^*) \geq \text{Tr}(X_{K_\gamma^*}^i Q) \geq \lambda_{\min}(Q) \|X_{K_\gamma^*}^i\|_2$. Applying Theorem \ref{Jbound} at optimality gives $\max_i J^i(K_\gamma^*) \leq \frac{c}{c-2} J_{\text{con},\gamma}^*$. Thus,
\begin{align}
   & \max_i \|X_{K_\gamma^*}^i\|_2 \leq \frac{c}{(c-2) \lambda_{\min}(Q)} J_{\text{con},\gamma}^*,\\
    & \max_i \|P_{K}^i\|_2 \leq \frac{c}{c-2} J_{\text{con}}(K \mid \gamma).
\end{align}
Substituting all of these explicit bounds back into the residual inequality \eqref{eq:Xkkstar},
\begin{align}
    \label{eq:residual_squared}
    \|\mathcal{X}(K, K_\gamma^*)\|_F^2 \leq &\left[ \left( \frac{2\sqrt{\gamma} \lambda_{N} \|B\|_2 J_{\text{con},\gamma}^*}{\lambda_{\min}(R)^{1/2} \lambda_{\min}(Q)^\frac{5}{2}} \right)^2 \left( \frac{c}{c-2} \right)^7  J_{\text{con}}(K \mid \gamma)^5 \right] \left( J_{\text{con}}(K \mid \gamma) - J_{\text{con},\gamma}^* \right).
\end{align}
Next, by the stated hypothesis, the current controller satisfies $J_{\text{con}}(K \mid \gamma) \leq M J_{\text{con},\gamma}^*$. Imposing this bound implies that if, for some $\epsilon > 0$, the static network parameters satisfy,
\begin{align}\label{eq:M5}
    &  M^5 \left( \frac{\gamma\lambda_{N} \|B\|_2}{\lambda_{\min}(R) \lambda_{\min}(Q)^5} \right) \left( \frac{c}{c-2} \right)^7 (J_{\text{con},\gamma}^*)^7   \leq \frac{1}{4}\left(\frac{\lambda_{\min}(R) \mu -1-\epsilon}{\mu(1+\epsilon^{-1})}\right),
\end{align}
then the squared residual bound strictly simplifies to,
\begin{equation}\label{eq:epsilonbound}
    \|\mathcal{X}(K, K_\gamma^*)\|_F^2 \leq \left(\frac{\lambda_{\min}(R) \mu -1-\epsilon}{\mu(1+\epsilon^{-1})}\right) \left( J_{\text{con}}(K \mid \gamma) - J_{\text{con},\gamma}^* \right).
\end{equation}

For any $\mu > \frac{1}{\lambda_{\min}(R)}$, selecting $\epsilon = \sqrt{\lambda_{\min}(R)\mu} - 1$, which maximizes the scalar factor $\frac{\lambda_{\min}(R) \mu - 1 - \epsilon}{\mu(1+\epsilon^{-1})}$ while ensuring $\epsilon > 0$, reduces condition \eqref{eq:M5} to,
\begin{align}\label{eq:M5final}
    &  M^5 \left( \frac{\gamma\lambda_{N} \|B\|_2}{\lambda_{\min}(R) \lambda_{\min}(Q)^5} \right) \left( \frac{c}{c-2} \right)^7 (J_{\text{con},\gamma}^*)^7   \leq \frac{\left(\sqrt{\lambda_{\min}(R) \mu}-1\right)^2}{4\mu}.
\end{align}
Substituting the residual bound \eqref{eq:epsilonbound} into the cost difference inequality in Lemma~\ref{lemma:subopt} yields,
\begin{align}
    &  J_{\text{con}}(K \mid \gamma) - J_{\text{con},\gamma}^* \leq \dfrac{1+\epsilon}{\lambda_{min}(R)} \|\nabla J_{\text{con}}(K \mid \gamma)\|_F^2   + \left( 1-\frac{1+\epsilon}{\mu \lambda_{\min}(R)}\right) \left( J_{\text{con}}(K \mid \gamma) - J_{\text{con},\gamma}^* \right),
\end{align}
which rearranges equivalently to $J_{\text{con}}(K \mid \gamma) - J_{\text{con},\gamma}^* \leq \mu \|\nabla J_{\text{con}}(K \mid \gamma)\|_F^2$. \hfill $\blacksquare$

The following algorithm iteratively walks the discount factor from the analytically guaranteed stable initialization $\gamma_0$ up to the unscaled system $\gamma = 1$, generating a jointly stabilizing controller.

\begin{algorithm}
\caption{Dynamic Consensus by Discount Annealing}
\label{alg:discount_annealing}
\begin{algorithmic}[1]
\Require System dynamics matrices $A, B$, communication modes $\{\lambda_2, \dots, \lambda_N\}$, and final optimization tolerance $\epsilon$.
\Ensure A jointly stabilizing consensus controller $K^*$ satisfying the final optimality gap $\epsilon$.

\State Initialize $K \gets 0$
\State \textbf{Initial Discounting:} Find the largest initial discount factor $\gamma \in (0, 1]$ satisfying:
\Statex \qquad $J_{\text{con}}(K \mid \gamma) \leq M n_x \lambda_{min}(Q)$.

\While{$\gamma < 1$}
    \State \textbf{Intermediate Policy Optimization:} 
    \Statex \qquad Run policy gradient starting from $K$ to find $K'$ such that:
    \Statex \qquad $J_{\text{con}}(K' \mid \gamma) - \inf_{K} J_{\text{con}}(K \mid \gamma) \leq n_x \lambda_{min}(Q)$
    
    \State \textbf{Update Controller:} $K \gets K'$
    
    \State \textbf{Discount Factor Inflation:} 
    \Statex \qquad Find $\gamma' \in [\gamma, 1]$ such that the cost is inflated by a factor between $2.5$ and $\frac{M}{2}$:
    \Statex \qquad $2.5 J_{\text{con}}(K \mid \gamma) \leq J_{\text{con}}(K \mid \gamma') \leq \frac{M}{2} J_{\text{con}}(K \mid \gamma)$
    
    \State \textbf{Update Discount Factor:} $\gamma \gets \gamma'$
\EndWhile

\State \textbf{Final Policy Optimization:} 
\Statex \qquad Run policy gradient starting from $K$ (at $\gamma = 1$) to find the final controller $K^*$ such that:
\Statex \qquad $J_{\text{con}}(K^*) - \inf_{K} J_{\text{con}}(K) \leq \epsilon$

\State \Return $K^*$
\end{algorithmic}
\end{algorithm}

\begin{theorem}\label{thm:alg_viability} Suppose that the network's spectral spread parameter satisfies the topological condition established in Theorem \ref{lem:grad_domination} for some $M>5$ and $\mu > \dfrac{1}{\lambda_{min}(R)}$.  Then the intermediate policy optimization subroutines in Algorithm \ref{alg:discount_annealing} are guaranteed to converge globally at a linear rate to the required intermediate tolerance. \end{theorem}
\textit{Proof:}
To show the convergence of the intermediate subroutines, it is sufficient to show that at the start of every intermediate policy optimization step , the active controller satisfies the strict bounded suboptimality invariant $J_{\text{con}}(K \mid \gamma) \leq M J_{\text{con},\gamma}^*$. Then according to Theorem \ref{lem:grad_domination} the gradient domination property ($J_{\text{con}}(K \mid \gamma) - J_{\text{con},\gamma}^* \leq \mu \Vert{}\nabla J_{\text{con}}(K \mid \gamma)\Vert{}_F^2$) holds uniformly, guaranteeing that the inner-loop policy gradient converges globally and at a linear rate to the required intermediate tolerance. We prove that this invariant holds at every iteration by induction. 

\textit{Base case:} At the beginning of the algorithm (Step 2), the initial discount factor $\gamma_0$ is explicitly selected such that $J_{\text{con}}(K=0 \mid \gamma_0) \leq M n_x \lambda_{\min}(Q)$. Because the LQR objective is strictly bounded below by the state penalty trace, the optimal cost satisfies $J_{\text{con},\gamma_0}^* \geq \text{Tr}(Q) \geq n_x \lambda_{\min}(Q)$. Substituting this lower bound yields,
$J_{\text{con}}(K=0 \mid \gamma_0) \leq M J_{\text{con},\gamma_0}^*.$
Thus, the gradient domination condition holds for the very first policy optimization step.

\textit{Inductive Step:} Assume that at iteration $k$ with discount factor $\gamma_k$, the policy optimization successfully returns a controller $K_k$ satisfying the intermediate convergence tolerance $\tau = n_x \lambda_{\min}(Q)$. Using the universal lower bound $J_{\text{con},\gamma_k}^* \geq \tau$, the cost of the optimized controller satisfies,
\begin{equation}J_{\text{con}}(K_k \mid \gamma_k) \leq J_{\text{con},\gamma_k}^* + \tau \leq 2 J_{\text{con},\gamma_k}^*\end{equation}
In the subsequent discount factor inflation step (Step 6), the algorithm dynamically selects a new discount factor $\gamma_{k+1}$ such that the cost is inflated by a factor of at most $\frac{M}{2}$. Therefore, evaluated at the new discount factor, the inherited controller $K_k$ satisfies,
\begin{equation}
J_{\text{con}}(K_k \mid \gamma_{k+1}) \leq \frac{M}{2} J_{\text{con}}(K_k \mid \gamma_k) \leq  M J_{\text{con},\gamma_k}^*
\end{equation}
Because the optimal consensus cost is a monotonically increasing function of the discount factor, we are guaranteed that $J_{\text{con},\gamma_k}^* \leq J_{\text{con},\gamma_{k+1}}^*$. Substituting this monotonic property yields, $J_{\text{con}}(K_k \mid \gamma_{k+1}) \leq M J_{\text{con},\gamma_{k+1}}^*.$ \hfill $\blacksquare$

According to Theorem \ref{thm:alg_viability} the condition $J_{\text{con}}(K \mid \gamma) \leq M J_{\text{con},\gamma}^*$ is strictly maintained at the initialization of every intermediate policy optimization subroutine. Because this satisfies the hypothesis of Theorem \ref{lem:grad_domination}, the objective landscape exhibits exact gradient domination ($J - J^* \leq M \Vert{}\nabla J\Vert{}_F^2$) uniformly throughout the algorithm's execution. This geometric property guarantees that standard policy gradient updates avoid spurious stationary points and achieve linear convergence to the global optimum at each phase, proving that the algorithmic requirements are fully realizable.

\begin{theorem}\label{thm:iterations}
Suppose that the tunable spectral spread bound in Assumption \ref{assumption:3} holds. Let $\gamma_0$ be the initial discount factor and $J_{\text{con}}^* = \inf_{K} J_{\text{con}}(K \mid \gamma=1)$. Then the algorithm is guaranteed to cease after at most,
$$ \left( \frac{4 c J_{\text{con}}^*}{(c-2) \lambda_{\min}(Q)} - 1 \right) \log\left(\frac{1}{\gamma_0}\right), $$
iterations. Furthermore, at the start of any iteration, it holds that $J_{\text{con}}(K \mid \gamma) \leq M \inf_{K'} J_{\text{con}}(K' \mid \gamma)$, and the final controller jointly stabilizes all active network.
\end{theorem}
\textit{Proof:}
Similar to the inductive step of proof of Theorem \ref{thm:alg_viability} $J_{\text{con}}(K \mid \gamma) \le 2 J_{\text{con},\gamma}^*$. From Theorem \ref{Jbound}, the maximum modal cost is strictly bounded by the network average scaled by $s=\frac{c}{c-2}$. Consequently,
\begin{align*}
    \max_{i} \|P^i_{K,\gamma}\|_2 &\le \max_{i} J(K, \lambda_i \mid \gamma) \le \frac{c}{c-2} J_{\text{con}}(K \mid \gamma) \le \frac{2c}{c-2} J_{\text{con}}^* \equiv 2s J_{\text{con},\gamma}^*.
\end{align*}
According to Theorem \ref{thm:gamma_inflation}, if the discount factor is inflated to $\bar{\gamma}$, the modal cost will at most double provided that,
\begin{equation}
    \dfrac{\bar{\gamma}}{\gamma} \leq \min_{i \in \{2, \dots, N\}} \left( 1 + \frac{\lambda_{\min}(Q)}{2 \left( \Vert P_{K,\gamma}^i \Vert_2 - \lambda_{\min}(Q) \right)} \right).
\end{equation}
The algorithm's update step enforces an updated discount factor $\gamma'$ that satisfies $\gamma' \ge \bar{\gamma}$ with,
\begin{align}
    \nonumber&  \frac{\gamma'}{\gamma} \ge 1 + \frac{\lambda_{\min}(Q)}{2 \left( \max_{i}\Vert P_{K,\gamma}^i \Vert_2 - \lambda_{\min}(Q) \right)}  \ge 1 + \frac{\lambda_{\min}(Q)}{4s J_{\text{con}}^* - 2\lambda_{\min}(Q)}.
\end{align}
The algorithm terminates once the discount factor reaches the undiscounted objective $\gamma = 1$. Let the minimum guaranteed step increment be defined as $q = \frac{\lambda_{\min}(Q)}{4s J_{\text{con}}^* - 2\lambda_{\min}(Q)}$. Because $2s J_{\text{con}}^* \geq \max_{i} \|P^i_{K,\gamma}\|_2 \geq \lambda_{\min}(Q)$, it is guaranteed that $q > 0$. Because the step ratio is at least $(1+q)$ at every iteration, the maximum possible number of iterations $N$ occurs under worst-case minimal growth, evaluated by solving $\gamma_0 (1+q)^N = 1$. Thus, the actual iteration count is upper-bounded by,
\begin{align*}
    N &\le \frac{\log(1/\gamma_0)}{\log(1+q)} \leq \left(1 + \frac{1}{q}\right) \log\left(\frac{1}{\gamma_0}\right) =\left( 1 + \frac{4s J_{\text{con}}^* - 2\lambda_{\min}(Q)}{\lambda_{\min}(Q)} \right) \log\left(\frac{1}{\gamma_0}\right) = \left( \frac{4 c J_{\text{con}}^*}{(c-2) \lambda_{\min}(Q)} - 1 \right) \log\left(\frac{1}{\gamma_0}\right).
\end{align*}\hfill $\blacksquare$

\section{Federated Learning Mechanism for Dynamic Consensus}\label{sec:federated}
In a connected network of $N$ agents, the graph Laplacian possesses $N$ eigenvalues comprising the zero consensus mode $\lambda_1 = 0$ and $N-1$ active non-zero disagreement eigenvalues ($\lambda_2, \dots, \lambda_N$). An exact $1$-to-$1$ mapping is induced across all $N$ nodes: the agent associated with $\lambda_1 = 0$ naturally serves as the central aggregator, while the remaining $N-1$ agents act as parallel client nodes. Assigning each client agent $i \in \{2, \dots, N\}$ responsibility for a single non-zero eigenvalue $\lambda_i$ directly exploits the modal decomposition of the consensus network. Furthermore, because the global consensus objective $J_{\text{con}}(K \mid \gamma) = \frac{1}{N-1} \sum_{i=2}^N J^i(K \mid \gamma)$ is a uniform average of decoupled modal costs, the expectation and gradient operators distribute linearly across modes, enabling the computational burden of zeroth-order trajectory generation to be parallelized across the clients.

Consider an unknown single-agent system governed by $x_i(k+1) = A x_i(k) + B u_i(k)$. To evaluate a consensus controller $K$ for the $i$-th network mode, the eigenvalue-scaled feedback law $u_i(k) = -\lambda_i K x_i(k)$ is applied, inducing the closed-loop modal dynamics $ x_i(k+1) = (A - \lambda_i B K) x_i(k)$.
Over a rollout horizon $\tau$, the resulting state trajectory $\{x_i(k)\}_{k=0}^{\tau}$ yields the empirical discounted modal cost,
\begin{equation}\label{eq:emp_modal_cost}
    \widehat{C}^i = \sum_{k=0}^{\tau} \gamma^k x_i(k)^\top \left( Q + \lambda_i^2 K^\top R K \right) x_i(k).
\end{equation}
To optimize the consensus policy without direct gradient access, a smoothed surrogate objective is formulated as $J_{\sigma,\text{con}}(K \mid \gamma) = \mathbb{E}_{U \sim \mathbb{S}_F(\sigma)} \left[ J_{\text{con}}(K + U \mid \gamma) \right]$, where $\sigma > 0$ denotes the smoothing radius and $\mathbb{S}_F(\sigma) \triangleq \{U \in \mathbb{R}^{n_u \times n_x} : \|U\|_F = \sigma\}$ represents the Frobenius sphere of radius $\sigma$. The smoothed consensus gradient decomposes across the active modes as,
\begin{equation}\label{eq:grad_decomp}
    \nabla J_{\sigma,\text{con}}(K \mid \gamma) = \frac{1}{N-1} \sum_{i=2}^N \nabla J_{\sigma}^i(K \mid \gamma).
\end{equation}
By Stokes' theorem, each modal gradient $\nabla J_{\sigma}^i(K \mid \gamma)$ can be estimated from finite-horizon rollouts along random perturbation directions \cite{fazel2018global, malik2019derivative}. Specifically, at optimization iteration $t$, the central aggregator broadcasts the current global policy $K_t$ to all client nodes. Each client $i \in \{2, \dots, N\}$ independently draws $n_s$ perturbation matrices $\{U_{i,s}\}_{s=1}^{n_s}$ uniformly from $\mathbb{S}_F(\sigma)$, executes local rollouts under the perturbed modal control law $u_i(k) = -\lambda_i (K_t +  U_{i,s}) x_i(k)$ to obtain empirical costs $\{\widehat{C}^i_s\}_{s=1}^{n_s}$, and constructs the local zeroth-order gradient estimator,
\begin{equation}\label{eq:local_zo_grad}
    \widehat{\nabla} J^i(K_t \mid \gamma) = \frac{1}{n_s} \sum_{s=1}^{n_s} \frac{n_x n_u}{\sigma^2} \widehat{C}^i_s U_{i,s}.
\end{equation}

The clients then transmit their local gradient estimates to the central aggregator, which computes the global policy gradient via federated averaging,
\begin{equation}\label{eq:fed_grad_agg}
    \widehat{\nabla} J_{\text{con}}(K_t \mid \gamma) = \frac{1}{N-1} \sum_{i=2}^N \widehat{\nabla} J^i(K_t \mid \gamma).
\end{equation}
While model-free optimization of the standard LQR problem using zeroth-order estimators is well-documented (e.g., \cite{fazel2018global, malik2019derivative}), the proposed framework extends derivative-free optimization to a sequence of LQR consensus objectives $J_{\text{con}}(K \mid \gamma)$. Specifically, the intermediate phases of the discount-annealing architecture in Algorithm~\ref{alg:discount_annealing} are mapped onto a federated model-free setup driven by zeroth-order gradient estimation, yielding the complete protocol summarized in Algorithm~\ref{alg:federated_annealing}.
\begin{theorem}\label{thm:federated}
Suppose the discount-annealing parameter $M$ and the network spectral-spread parameter $c$ satisfy the topological threshold~\eqref{eq:topology_condition}. For a sufficiently small smoothing radius $\sigma > 0$, a sufficiently large number of local samples $n_s$, a sufficiently long rollout horizon $\tau$, and a step size satisfying $0< \eta \leq (1/2L_\gamma)$, Algorithm~\ref{alg:federated_annealing} converges with high probability to the same $\epsilon$-optimal global consensus controller $K^*$ as the centralized model-based scheme in Algorithm~\ref{alg:discount_annealing}.
\end{theorem}
\textit{Proof:} The proof is established by analyzing four interconnected properties of Algorithm~\ref{alg:federated_annealing}: (i) unbiasedness and error decomposition of the federated zeroth-order estimator, (ii) step-size selection and sublevel-set invariance, (iii) global convergence via gradient dominance.

\emph{(i) Unbiasedness and error decomposition of the federated zeroth-order estimator:} 
For any discount factor $\gamma \in [\gamma_0, 1]$ and any policy $K_t$ within the stabilizing domain, let $J_{\sigma}^i(K_t \mid \gamma) \triangleq \mathbb{E}_{U_{i,s} \sim \mathbb{S}_F(\sigma)}[J^i(K_t + U_{i,s} \mid \gamma)]$ denote the smoothed modal objective over the Frobenius sphere of radius $\sigma$.  According to \cite{fazel2018global} the exact gradient of the smoothed modal cost satisfies,
\begin{equation}
    \nabla J_{\sigma}^i(K_t \mid \gamma) = \mathbb{E}_{U_{i,s} \sim \mathbb{S}_F(\sigma)}\!\left[\frac{n_x n_u}{\sigma^2} J^i(K_t + U_{i,s} \mid \gamma) U_{i,s}\right].
\end{equation}
Because the global consensus objective decomposes linearly across the $N-1$ active disagreement modes, synchronous gradient aggregation at the central server yields, $\mathbb{E}\!\left[\widehat{\nabla} J_{\text{con}, t}\right] = \frac{1}{N-1} \sum_{i=2}^N \mathbb{E}\!\left[\widehat{\nabla} J^i_t\right],$
where the total estimation error can be represented by $E_t \triangleq \widehat{\nabla} J_{\text{con}, t} - \nabla J_{\text{con}}(K_t \mid \gamma)$. Similar to \cite{fazel2018global}
$\|E_t\|$ can be made arbitrarily small with high probability by choosing $\sigma$ sufficiently small and $(n_s, \tau)$ sufficiently large.

\emph{(ii) Step-size selection and sublevel-set invariance:}
Define the discounted stabilizing sublevel set $\mathcal{S}_\gamma(M) \triangleq \{K : J_{\text{con}}(K \mid \gamma) \leq M J_{\text{con},\gamma}^*\}$. On $\mathcal{S}_\gamma(M)$, the consensus cost $J_{\text{con}}(\cdot \mid \gamma)$ is locally $L_\gamma$-smooth and coercive as an extension of the single-agent LQR \cite{hu2023toward}. Unlike local-averaging schemes that suffer from client drift across heterogeneous modal dynamics, Algorithm~\ref{alg:federated_annealing} aggregates local gradients prior to executing a single global update $K_{t+1} = K_t - \eta \widehat{\nabla} J_{\text{con}, t}$. Consequently, the global iterate evolves directly along a perturbed first-order descent trajectory of the centralized objective. Similar to the argument in \cite{hu2023toward}, for $L_\gamma$-smooth functions with a step size $\eta \leq \frac{1}{2L_\gamma}$,
\begin{align}
    J_{\text{con}}(K_{t+1} \mid \gamma) &\leq J_{\text{con}}(K_t \mid \gamma) - \eta \left\langle \nabla J_{\text{con}}(K_t \mid \gamma), \nabla J_{\text{con}}(K_t \mid \gamma) + E_t \right\rangle + \frac{L_\gamma \eta^2}{2} \left\|\nabla J_{\text{con}}(K_t \mid \gamma) + E_t\right\|_F^2 \nonumber \\
    &\leq J_{\text{con}}(K_t \mid \gamma) - \frac{\eta}{2} \|\nabla J_{\text{con}}(K_t \mid \gamma)\|_F^2 + \frac{\eta}{2} \|E_t\|_F^2. \label{eq:perturbed_descent}
\end{align}
Whenever $K_t$ lies outside the target suboptimality neighborhood, choosing $\sigma$ sufficiently small and $(n_s, \tau)$ sufficiently large ensures $\|E_t\|_F < \|\nabla J_{\text{con}}(K_t \mid \gamma)\|_F$ with high probability, guaranteeing monotonic cost decrease $J_{\text{con}}(K_{t+1} \mid \gamma) \leq J_{\text{con}}(K_t \mid \gamma) \leq M J_{\text{con},\gamma}^*$ and preserving invariance of $\mathcal{S}_\gamma(M)$ across all inner iterations.

\emph{(iii) Gradient dominance and global convergence:}
Because $M$ and $c$ satisfy~\eqref{eq:topology_condition}, Theorem~\ref{lem:grad_domination} establishes that $J_{\text{con}}(K \mid \gamma)$ satisfies the $\mu$-gradient dominance condition $J_{\text{con}}(K \mid \gamma) - J_{\text{con},\gamma}^* \leq \mu \|\nabla J_{\text{con}}(K \mid \gamma)\|_F^2$ throughout $\mathcal{S}_\gamma(M)$ for every $\gamma \in [\gamma_0, 1]$, ensuring that no spurious local minima exist within the invariant sublevel set. Defining the suboptimality gap $\Delta_t \triangleq J_{\text{con}}(K_t \mid \gamma) - J_{\text{con},\gamma}^*$ and substituting the gradient dominance inequality into~\eqref{eq:perturbed_descent} yields the one-step contraction,
\begin{equation}
    \Delta_{t+1} \leq \left(1 - \frac{\eta}{2\mu}\right)\Delta_t + \frac{\eta}{2} \|E_t\|_F^2.
\end{equation}
Suppose $\|E_t\|_F \leq E_{\max}$ across $t = 0, \dots, T-1$. Unrolling this recursion and bounding the resulting geometric sum via $\sum_{k=0}^{T-1}(1 - \frac{\eta}{2\mu})^k < \frac{2\mu}{\eta}$ gives,
\begin{equation}
    \Delta_T \leq \left(1 - \frac{\eta}{2\mu}\right)^T \Delta_0 + \mu E_{\max}^2 \leq \exp\!\left(-\frac{\eta}{2\mu}T\right) M J_{\text{con},\gamma}^* + \mu E_{\max}^2.
\end{equation}
For any target accuracy $\delta_{\text{target}} > 0$, selecting $\sigma$ sufficiently small and $(n_s, \tau)$ sufficiently large guarantees $\mu E_{\max}^2 \leq \frac{1}{2}\delta_{\text{target}}$ with high probability, so that $\Delta_T \leq \delta_{\text{target}}$ holds within $T = \mathcal{O}\!\left(\frac{\mu}{\eta} \log \frac{M J_{\text{con},\gamma}^*}{\delta_{\text{target}}}\right)$ iterations. Setting $\delta_{\text{target}} = n_x \lambda_{\min}(Q)$ establishes the iteration bound for each intermediate subroutine, while setting $\delta_{\text{target}} = \epsilon$ at $\gamma = 1$ guarantees linear high-probability convergence to an $\epsilon$-optimal neighborhood of the unique global consensus controller $K^*$, matching Algorithm~\ref{alg:discount_annealing}. \hfill $\blacksquare$

\begin{algorithm}
\caption{Federated Model-Free Consensus Algorithm}
\label{alg:federated_annealing}
\begin{algorithmic}[1]
\Require Discount Annealing parameter $M$, ZO parameters $(n_s, \tau, \sigma)$, step size $\eta$.
\Ensure Global consensus controller $K^*$.

\State \textbf{Server:} Initialize global policy $K \gets 0$
\State \textbf{Server:} Find initial discount factor $\gamma_0 \in (0, 1]$ satisfying $J_{\text{con}}(K \mid \gamma) \leq M n_x \lambda_{\min}(Q)$ via a federated forward-pass.
\While{$\gamma < 1$}
    \State \textbf{Federated Inner-Loop Optimization:} 
    \State Set $K_0 = K$.
    \For{$t = 0, 1, \dots, T-1$}
        \State \textbf{Server:} Broadcast $K_t$ to all agents.
        \State \textbf{Agents (in parallel for all $i$):}
        \For{$s = 1, \dots, n_s$}
            \State Sample $U_{i,s} \sim \mathbb{S}_F(\sigma)$
            \State set $\widehat{K}_{i,s} = K_t + U_{i,s}$
            \State simulate agent $i$ for $\tau$ steps with $\lambda_i \widehat{K}_{i,s}$.
            \State Compute local modal cost $\widehat{C}^i_s$.
        \EndFor
        \State Let $\widehat{\nabla} J^i_t = \frac{1}{n_s} \sum_{s=1}^{n_s} \frac{n_x n_u}{\sigma^2} \widehat{C}^i_s U_{i,s}$
        \State Transmit $\widehat{\nabla} J^i_t$ to Server.  
        \State \textbf{Server:} Set: $\widehat{\nabla} J_{\text{con}, t} = \frac{1}{N-1} \sum_{i=2}^N \widehat{\nabla} J^i_t$
        \State \textbf{Server:} Update: $K_{t+1} = K_t - \eta \widehat{\nabla} J_{\text{con}, t}$
        \State \textbf{Server:} \parbox[t]{0.7\linewidth}{Break if empirical optimization gap falls below $n_x \lambda_{min}(Q)$.\strut}
    \EndFor
    \State \textbf{Server:} Update baseline controller $K \gets K_t$
    
    \State \textbf{Federated Discount Inflation:} 
    \State \textbf{Server:} Initialize bounds $\gamma_{\text{low}} = \gamma$ and $\gamma_{\text{high}} = 1$.
    \Loop
        \State \textbf{Server:} \parbox[t]{0.7\linewidth}{Broadcast $\gamma_{\text{test}} = \frac{1}{2}(\gamma_{\text{low}} + \gamma_{\text{high}})$ and updated policy $K$ to all agents.\strut}        
        \State \textbf{Agents (in parallel for all $i$):}
        \State Simulate for $\tau$ steps with $u = -\lambda_i K x$.
        \State Transmit $\widehat{J}^i_{\text{test}}$ to Server.
        
        \State \textbf{Server:} Aggregate: $\widehat{J}_{\text{con}}' = \frac{1}{N-1} \sum_{i=2}^N \widehat{J}^i_{\text{test}}$
        
        \State \textbf{Server:} Evaluate inflation criteria:
        \If{$2.5 \widehat{J}_{\text{con}} \leq \widehat{J}_{\text{con}}' \leq \frac{M}{2} \widehat{J}_{\text{con}}$}
            \State $\gamma \gets \gamma_{\text{test}}$ and \textbf{break loop}.
        \ElsIf{$\widehat{J}_{\text{con}}' > \frac{M}{2} \widehat{J}_{\text{con}}$}
            \State $\gamma_{\text{high}} = \gamma_{\text{test}}$ 
        \Else
            \State $\gamma_{\text{low}} = \gamma_{\text{test}}$ 
        \EndIf
    \EndLoop
\EndWhile

\State \textbf{Final Optimization:} Execute lines 5-16 at $\gamma = 1$ until final tolerance $\epsilon$ is met.
\State \Return $K^*$
\end{algorithmic}
\end{algorithm}

\section{Simulation Results}
\begin{figure}[htbp]
    \centering
    \begin{minipage}[b]{0.45\textwidth}
        \centering
        \includegraphics[width=\linewidth]{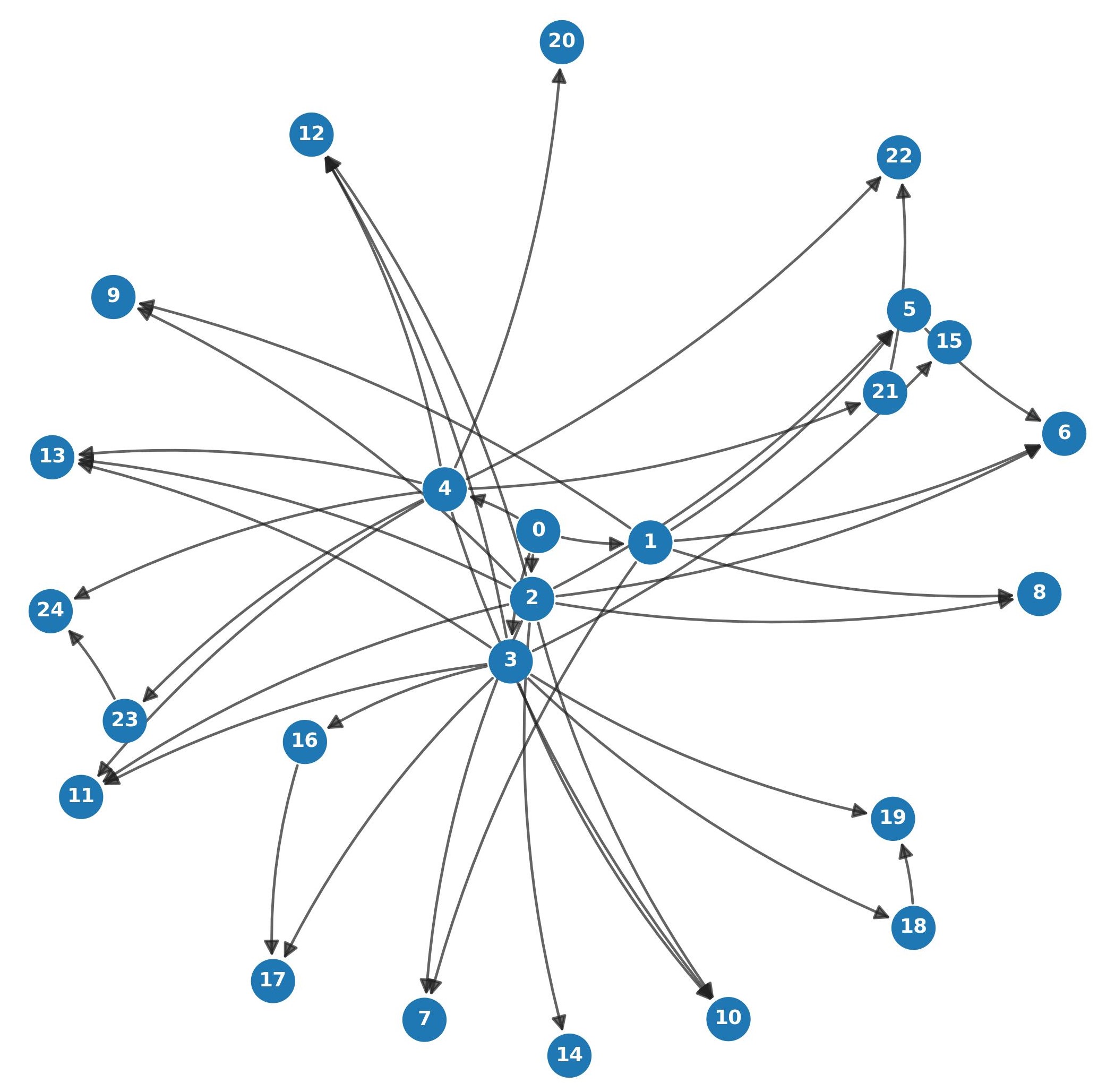}
        \caption{The communication graph in Example 1.}
        \label{fig:topology_Example1}
    \end{minipage}
    \hfill
    \begin{minipage}[b]{0.45\textwidth}
        \centering
        \includegraphics[width=\linewidth]{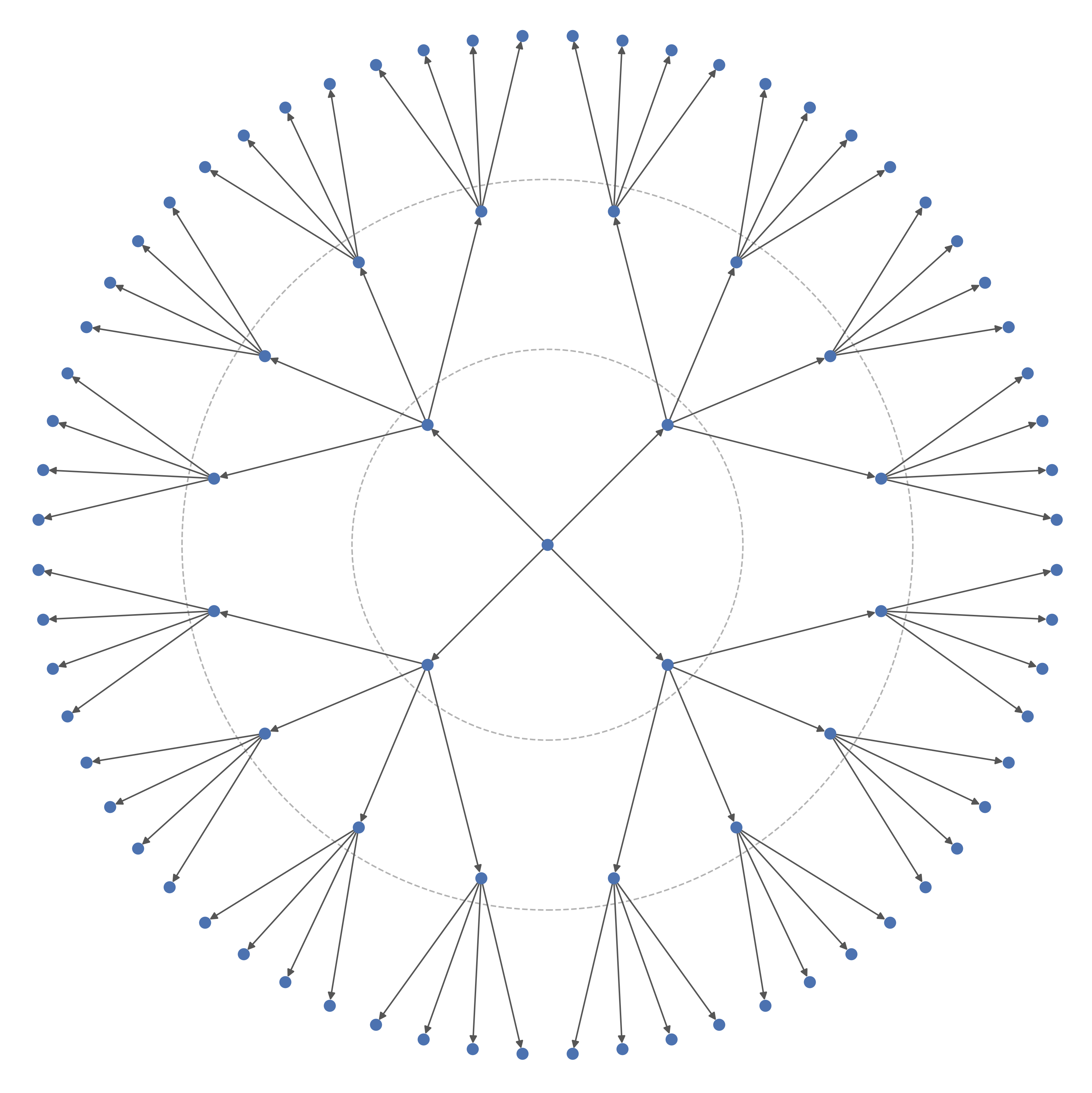}
        \caption{The communication graph in Example 3}
        \label{fig:hierarchical}
    \end{minipage}
\end{figure}

\textbf{Example 1: Federated policy gradient for consensus in a 25-node network} The multi-agent system consists of $N = 25$ identical linear discrete-time subsystems with a communication graph as given in Figure \ref{fig:topology_Example1}. The open-loop system matrices are defined as:
\begin{equation}
    A = \begin{bmatrix} 
    1.2 & 1.0 & 0.2 & -0.1 \\ 
    0.0 & 1.05 & 0.1 & 0.3 \\ 
    0.0 & 0.0 & 0.8 & 0.5 \\ 
    0.0 & 0.0 & -0.5 & 0.8 
    \end{bmatrix}, \quad 
    B = \begin{bmatrix} 
    0.0 & 0.1 \\ 
    1.0 & 0.0 \\ 
    0.0 & 1.0 \\ 
    0.5 & 0.2 
    \end{bmatrix}
\end{equation}

Let $Q = I_4$, $R = I_2$, and $\gamma_0=0.1$, $\eta = 10^{-3}$. The federated policy gradient algorithm was executed over 20 independent trials to account for the stochastic nature of the zeroth-order estimator. The hyper-parameters were set to $\tau = 1000$ simulation steps, batch size $n_s = 100$, and $\sigma = 0.1$.

\begin{figure}[htbp]
    \centering
    \includegraphics[width=0.8\textwidth]{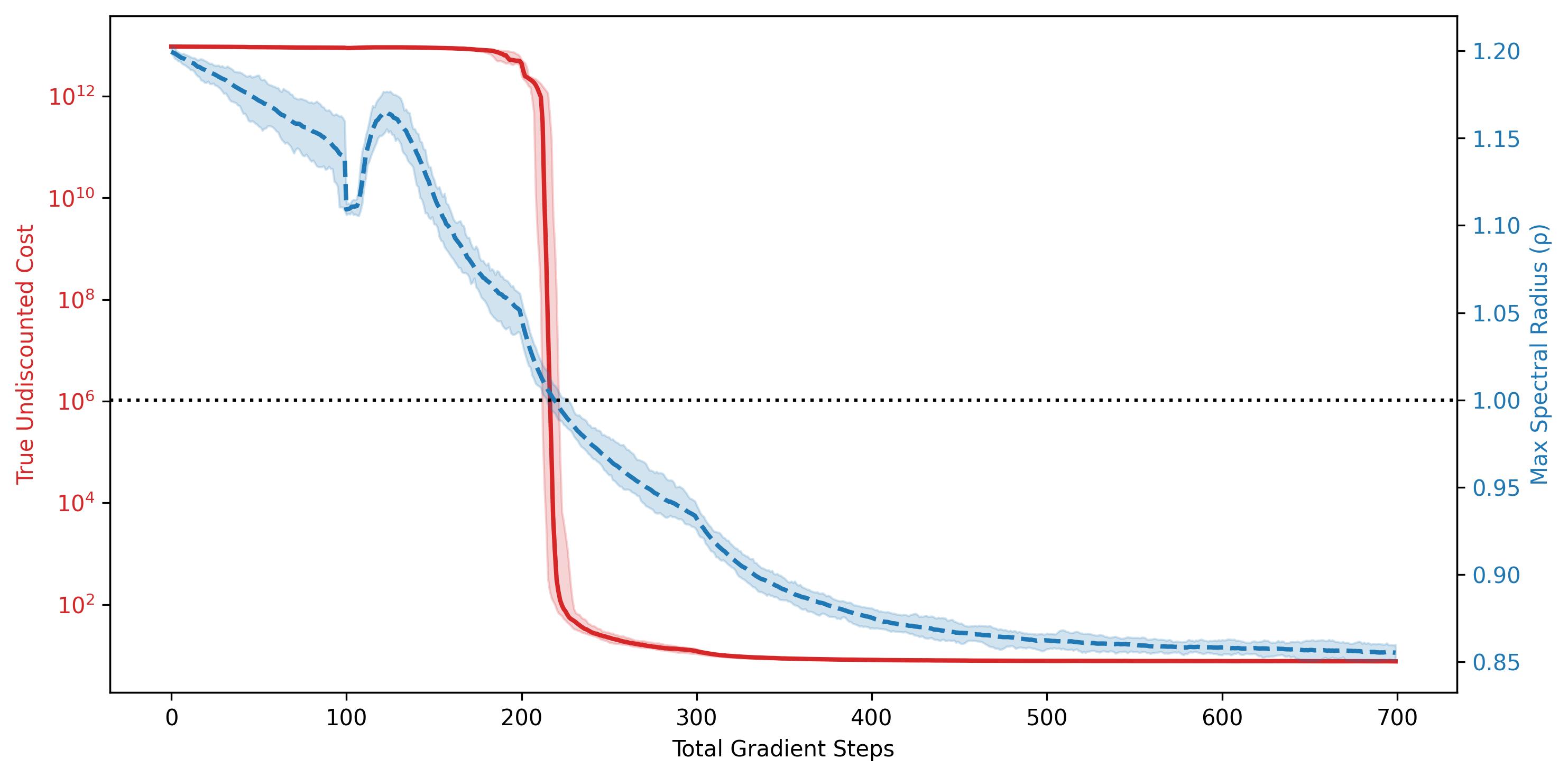}
    \caption{The undiscounted cost (red, log scale) and the maximum spectral radius (blue) over the total gradient steps in Example 1 over 20 trials.}
    \label{fig:learning_curve}
\end{figure}

The simulation results demonstrate the successful convergence and consensus capabilities of the proposed federated model-free learning algorithm. As depicted in the learning curve in Figure \ref{fig:learning_curve}, the system initializes in a highly unstable state, characterized by exceptionally high true undiscounted cost. 
To accurately evaluate the algorithm's robustness across all 20 independent trials, the central trajectories plot the median true cost and the mean spectral radius, while the shaded regions strictly represents the full empirical variance observed at each gradient step.
Through the progressive federation, the algorithm systematically updates the controller, leading to a critical phase transition near 220 gradient steps where the spectral radius definitively crosses the stability boundary ($\rho < 1$) and the true cost precipitously drops by over twelve orders of magnitude. The execution logs further illustrate this progression, showing how the algorithm stably transitions from $\gamma = 0.1$ up to the undiscounted setting ($\gamma = 1.0$). Finally, the theoretical verification confirms strict closed-loop modal stability across all active network modes. The worst-case spectral radius is safely constrained to 0.8555, perfectly matching the converged empirical trajectory and proving the high reliability and tight variance of the approach across all 20 independent trials.

\begin{figure}[htbp]
    \centering
    \includegraphics[width=.98\textwidth]{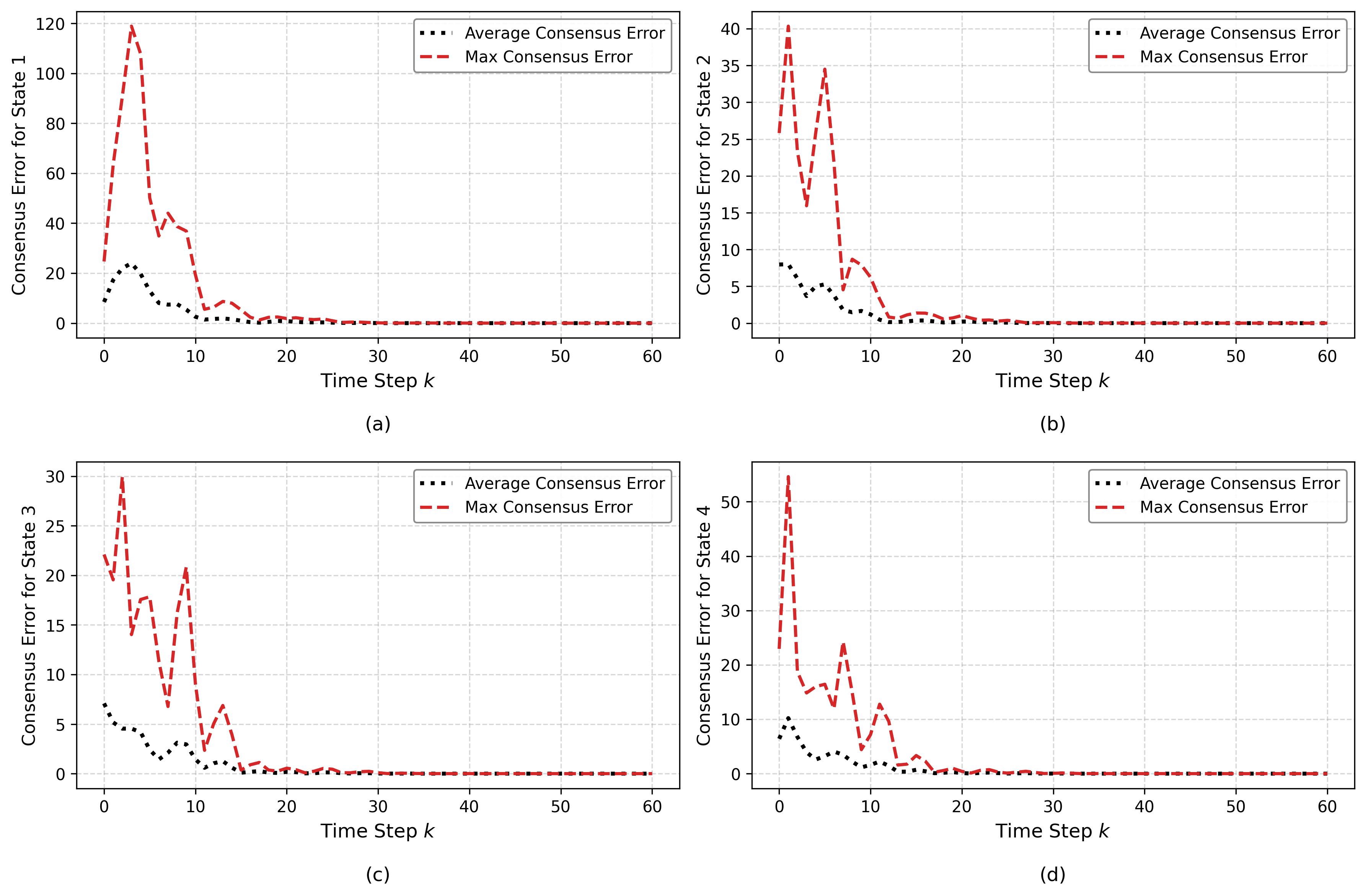}
    \caption{Dynamic consensus of 25 agents in Example 1 with maximum and average consensus error for (a) State
1 of all the agents, (b) State 2 of all the agents, (c) State 3 of all the
agents, and (d) State 4 of all the agents.
.}
    \label{fig:trajectories}
\end{figure}

Using the final optimal controller $K^*$, a 60-step time-domain simulation was performed on the full 25-node network. Initial states were randomly sampled from a uniform distribution bounded by $[0, 10]$. Figure \ref{fig:trajectories} demonstrates that all 25 agents successfully achieve consensus. This is the merit of federated learning while the non-federated schemes in \cite{feng2021consensusability,feng2021q} and the data-driven method in \cite{TCNS2026} all become infeasible.  In contrast, The federated scheme consistently achieves feasibility across all trials.

\textbf{Example 2: Comparison of feasibility} To evaluate the statistical feasibility of the proposed approach against existing literature, we consider a highly cross-coupled, unstable discrete-time linear system. The open-loop state matrix $A$ and the input matrix $B$ are given by,
\begin{equation}
    A = \begin{bmatrix}
     1.1 &  0.2 &  0.0 & -0.3 \\
     0.0 &  0.9 &  0.4 &  0.0 \\
    -0.2 &  0.0 &  1.2 &  0.5 \\
     0.0 & -0.3 &  0.0 &  1.0
    \end{bmatrix}, \quad
    B = \begin{bmatrix}
     0.2 & -0.1 \\
     1.0 &  0.2 \\
     0.1 &  0.8 \\
    -0.2 &  1.0
    \end{bmatrix}.
\end{equation}

Let $Q=I_4$ and $R=I_2$. To test the feasibility of different approaches in finding a consensus policy 10 connected, undirected communication topologies were generated and depicted in Table \ref{tab:feasibility_results}.

The feasibility of finding a stabilizing coupling gain was evaluated for the non-federated schemes in \cite{feng2021consensusability,feng2021q} and the data-driven method in \cite{TCNS2026} and also for our federated method.
The feasibility of the methods for each communication graph, are summarized in Table \ref{tab:feasibility_results}.
As shown in Table \ref{tab:feasibility_results}, the Q-learning method without a federation mechanism in \cite{feng2021q} and its model-based equivalence in \cite{feng2021consensusability} fails to find a feasible solution for most communication graphs, primarily due to its conservative reliance on separation of the coupling gain and the Riccati-based policy. This conservatism causes the method to become infeasible when the algebraic connectivity is low or the eigenvalue spread is wide. The data-driven approach in \cite{TCNS2026} shows more flexibility but yet remains ineffective for six communication graphs. In contrast, even in these relatively simple, low-dimensional dynamic settings and straightforward communication topologies, the effectiveness of our proposed federated method is highly evident. It consistently overcomes the conservativeness of previous approaches, guaranteeing a feasible solution across all tested network configurations.
\begin{table*}[htbp]
    \centering
    \caption{Feasibility Comparison of Control Methods Across Communication Topologies}
    \label{tab:feasibility_results}
    \renewcommand{\arraystretch}{1.8} 
    \begin{tabular}{|l|c|c|c|c|c|}
        \hline
        {Topology} & 
        $\vcenter{\hbox{\includegraphics[width=0.11\textwidth]{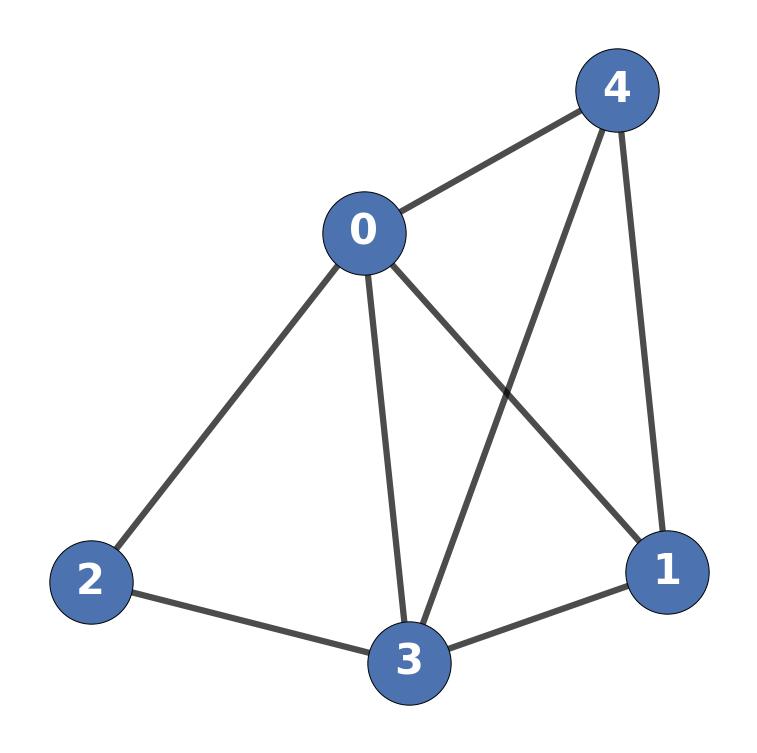}}}$ & 
        $\vcenter{\hbox{\includegraphics[width=0.11\textwidth]{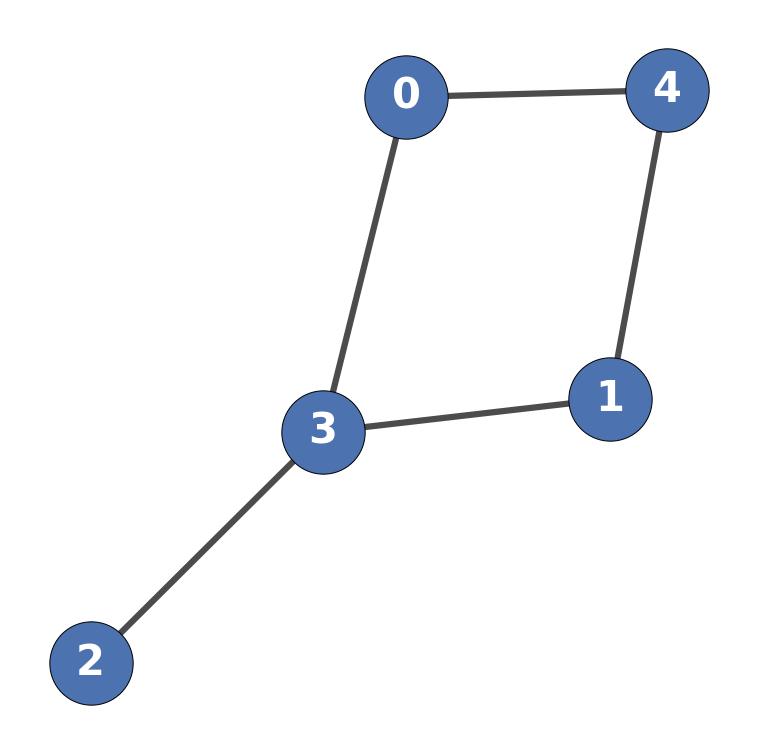}}}$ & 
        $\vcenter{\hbox{\includegraphics[width=0.11\textwidth]{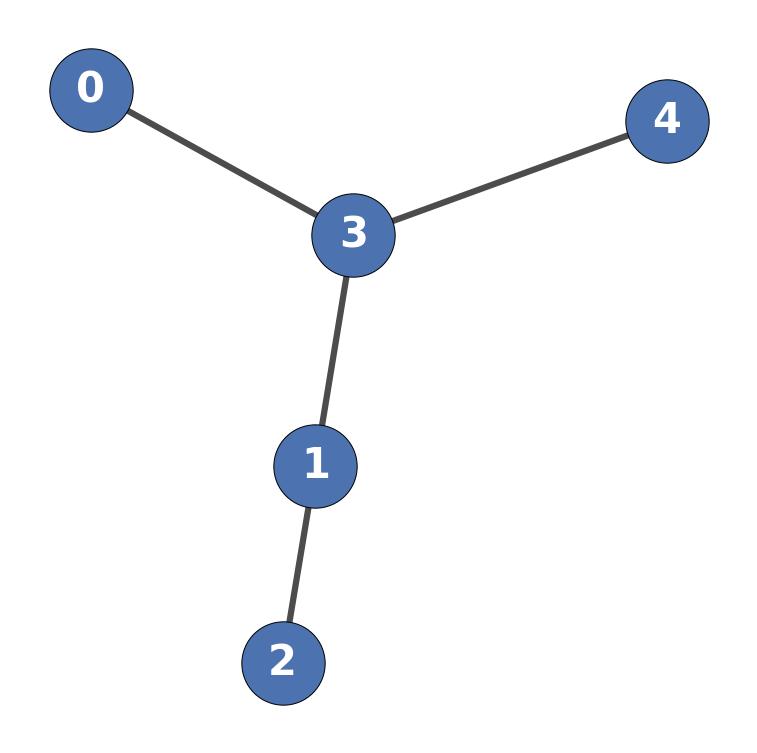}}}$ & 
        $\vcenter{\hbox{\includegraphics[width=0.11\textwidth]{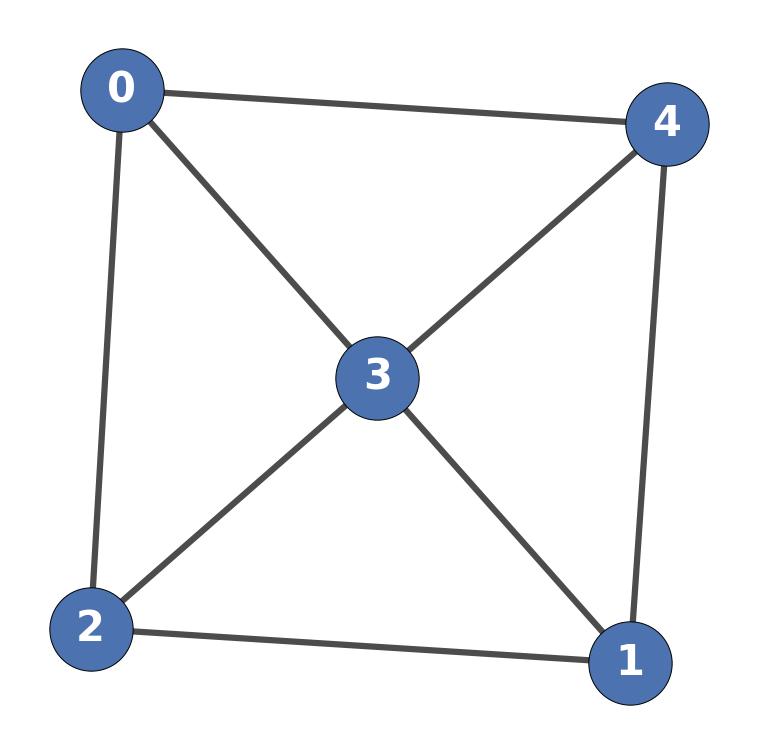}}}$ & 
        $\vcenter{\hbox{\includegraphics[width=0.11\textwidth]{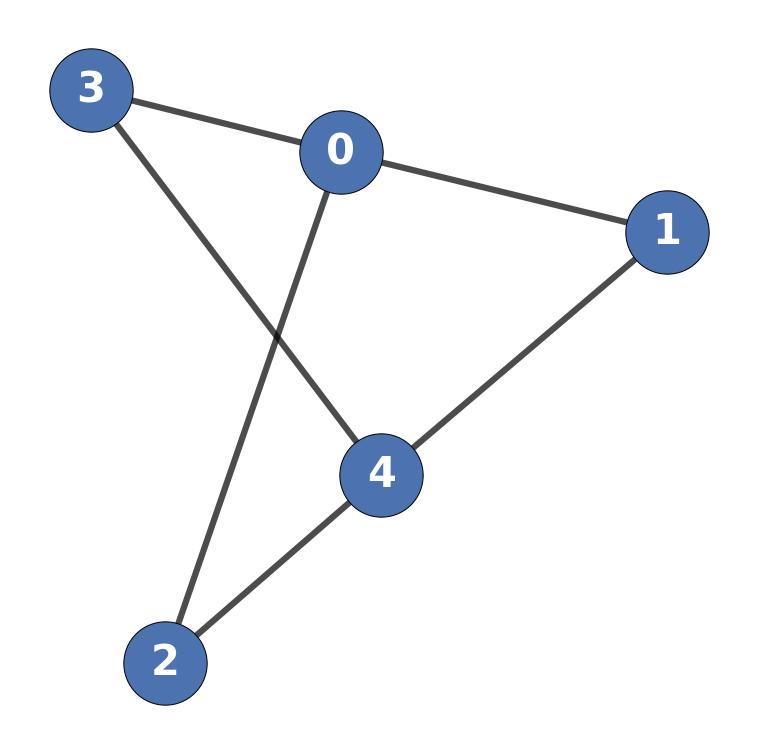}}}$ \\
        \hline
        \cite{TCNS2026} & \textbf{Feasible} & Infeasible & Infeasible & \textbf{Feasible} & \textbf{Feasible} \\
        \hline
        \cite{feng2021consensusability,feng2021q}      & Infeasible & Infeasible & Infeasible & \textbf{Feasible} & Infeasible \\
        \hline
        Federated Consensus & \textbf{Feasible} & \textbf{Feasible} & \textbf{Feasible} & \textbf{Feasible} & \textbf{Feasible} \\
        \hline
        \noalign{\vspace{2mm}}
        \hline
        Topology & 
        $\vcenter{\hbox{\includegraphics[width=0.11\textwidth]{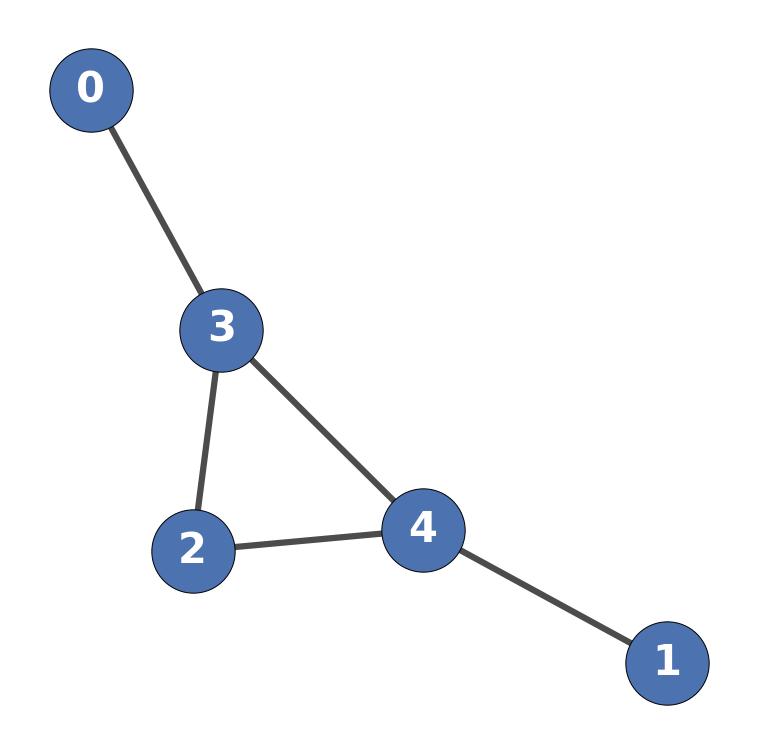}}}$ & 
        $\vcenter{\hbox{\includegraphics[width=0.11\textwidth]{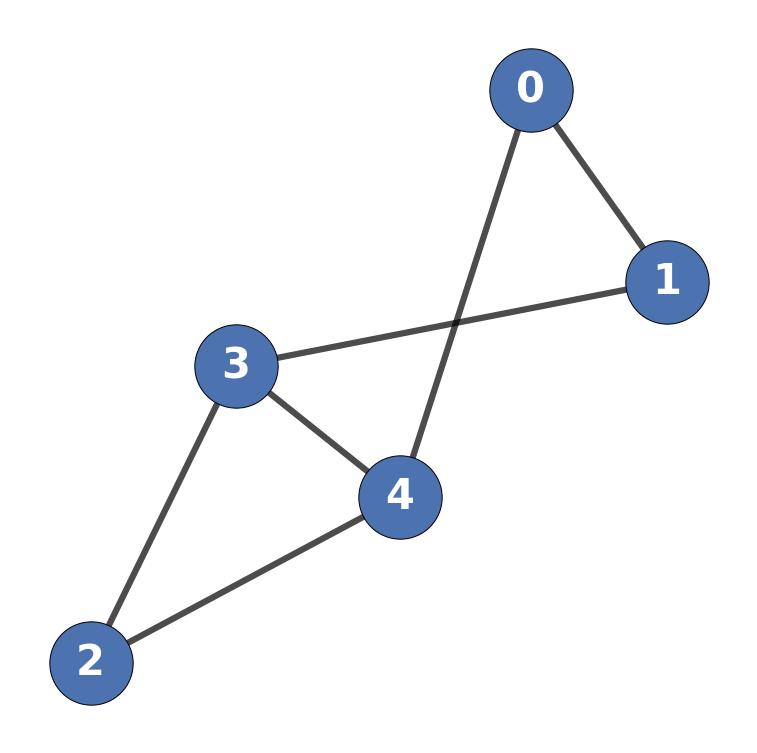}}}$ & 
        $\vcenter{\hbox{\includegraphics[width=0.11\textwidth]{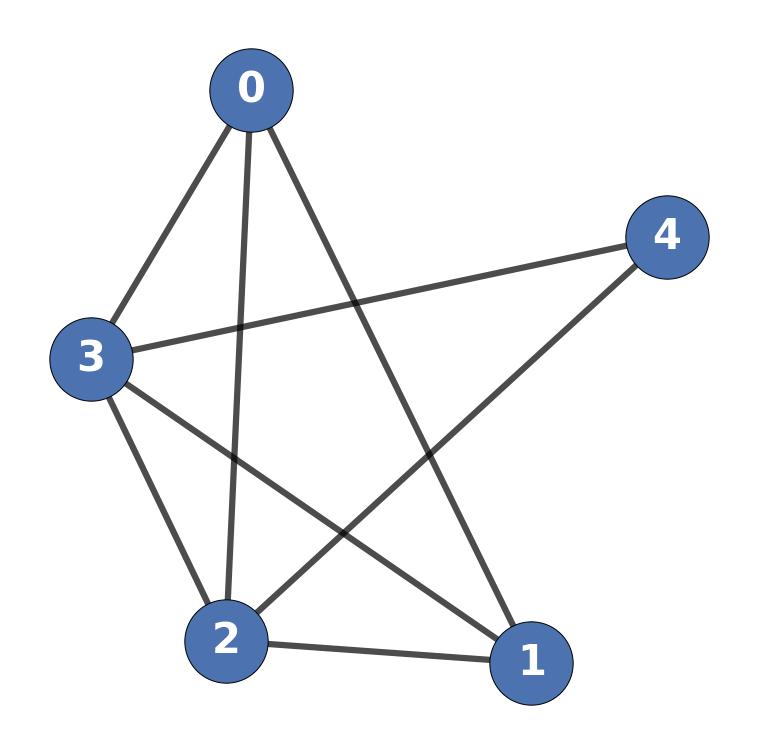}}}$ & 
        $\vcenter{\hbox{\includegraphics[width=0.11\textwidth]{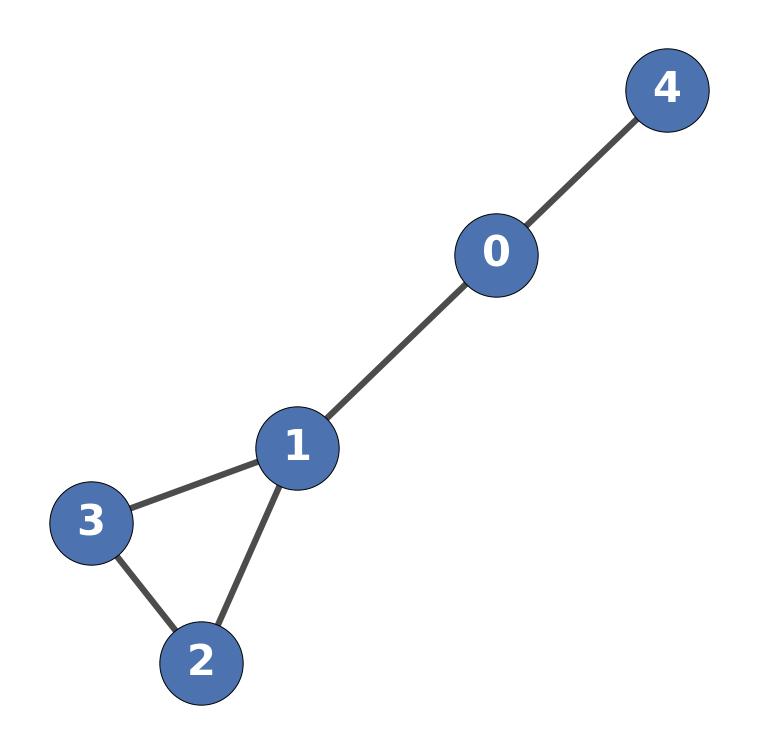}}}$ & 
        $\vcenter{\hbox{\includegraphics[width=0.11\textwidth]{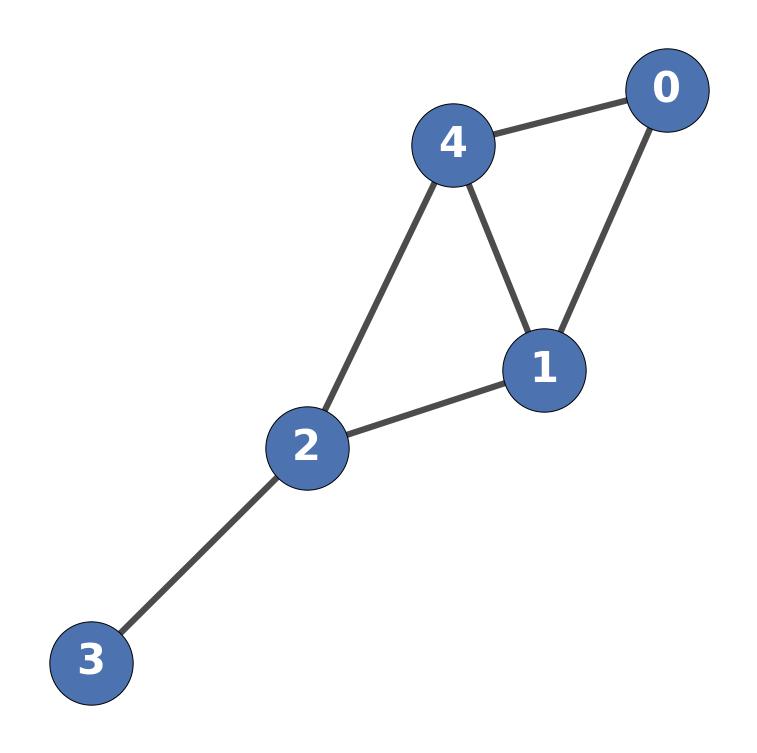}}}$ \\
        \hline
        \cite{TCNS2026} & Infeasible & Infeasible & \textbf{Feasible} & Infeasible & Infeasible \\
        \hline
        \cite{feng2021consensusability,feng2021q}      & Infeasible & Infeasible & Infeasible & Infeasible & Infeasible \\
        \hline
        Federated Consensus & \textbf{Feasible} & \textbf{Feasible} & \textbf{Feasible} & \textbf{Feasible} & \textbf{Feasible} \\
        \hline
    \end{tabular}
\end{table*}

\subsection*{Example 3: Dynamic consensus in the presence of heterogeneity and varying number of agents} 
This example models a heterogeneous multi-agent consensus problem over a hierarchical communication graph as shown in Figure \ref{fig:hierarchical}. The three evaluation cases, comprising $N=5$, $N=21$, and $N=85$ agents, correspond directly to increasing the depth of this hierarchical network to two, three, and four  layers, respectively.
The baseline nominal system for the network is defined by the following state and input matrices,
$$A_{nom} = \begin{bmatrix} 1.0 & 4.0 & -0.7 \\ 0.0 & 1.0 & 0.0 \\ 0.0 & 0.0 & 1.0 \end{bmatrix}, \quad B_{nom} = \begin{bmatrix} 0.0 & 1.0 \\ 2.0 & 0.0 \\ 0.0 & 3.0 \end{bmatrix}$$

\begin{figure}[htbp]
    \centering
    \includegraphics[width=0.8\textwidth]{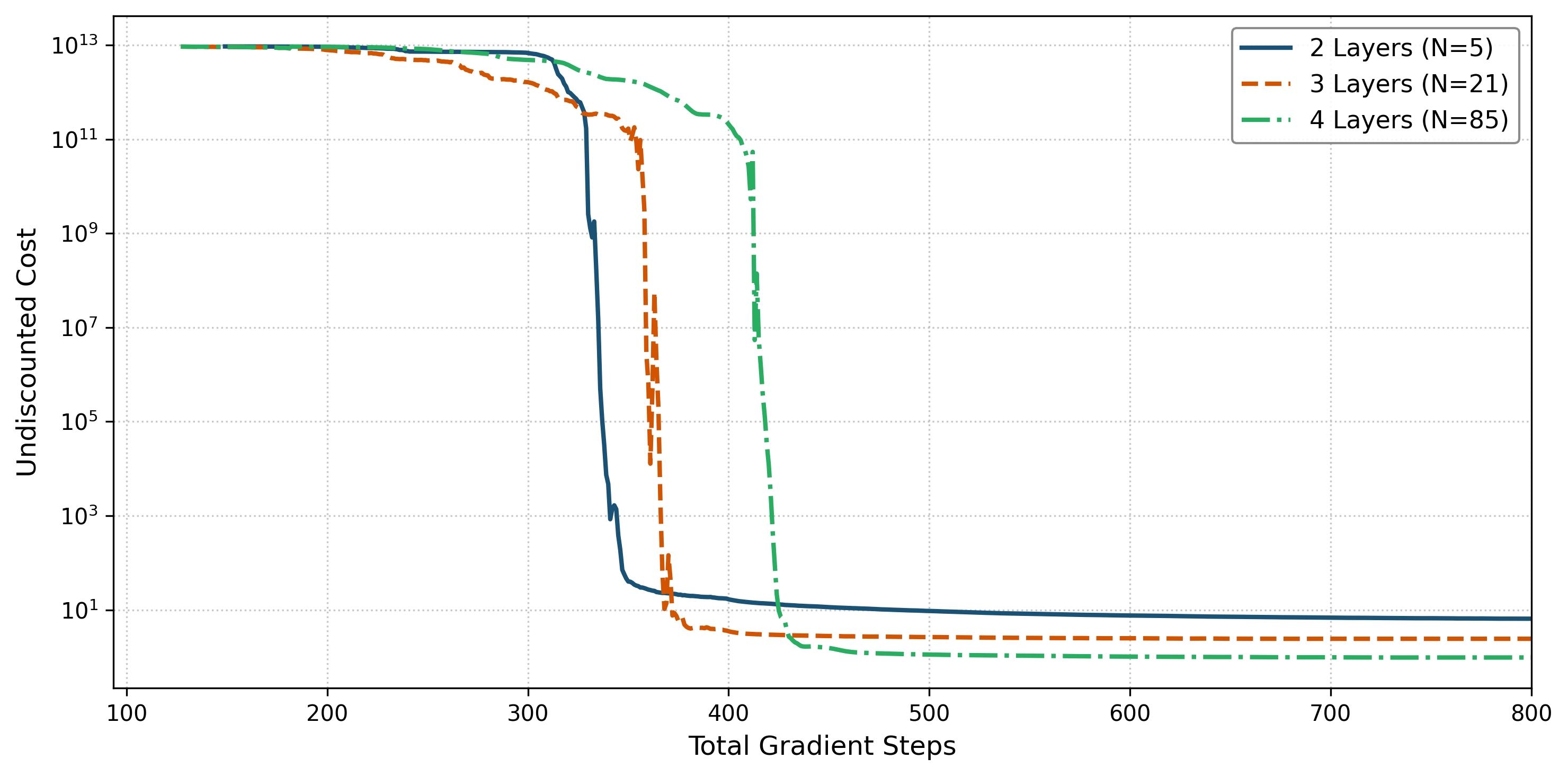}
    \caption{Evolution of the true undiscounted cost over total gradient steps for network sizes $N \in \{5, 21, 85\}$.}
    \label{fig:Jcomparison}
\end{figure}

To model heterogeneity, each agent $i$ receives a unique set of parameter perturbations. Specifically,  uncertainties drawn from a normal distribution $\mathcal{N}(0, 1)$ are injected into the off-diagonal elements of the nominal matrices. The federated zeroth-order learning algorithm successfully computes a robust global control policy. The algorithm trains the controller over the active network modes (the non-zero eigenvalues of the Laplacian matrix) using a discount-inflation bisection method. This allows the system to gradually discover a stabilizing policy without requiring explicit knowledge of the mismatched agent matrices. Figure \ref{fig:Jcomparison} illustrates the evolution of the true undiscounted cost, plotted on a logarithmic scale, as a function of the total gradient steps for varying network sizes ($N \in \{5, 21, 85\}$). This figure suggests that despite the pronounced delay in initial convergence, larger network sizes, such as $N=85$, are capable of achieving a slightly lower asymptotic cost floor compared to their smaller counterparts.

\section{Conclusion}
This paper addressed the dynamic consensus problem based on modal decomposition through the lens of federated optimization. An exact, parameterized suboptimality bound is established that isolates the aggregate policy gradient from the network covariance mismatch. Based on this formulation, it is shown that under a specific Laplacian spectral threshold, the consensus problem satisfies gradient dominance condition, thereby precluding spurious stationary points and mathematically guaranteeing convergence to the global optimum. Extensive numerical simulations validated these theoretical guarantees across diverse network topologies, sizes, and dynamic conditions.  Comparative feasibility analyses across multiple unweighted graphs demonstrated that the federated policy gradient approach consistently avoids the conservativeness of classical separation-based techniques, guaranteeing stabilizing solutions where existing model-based and data-driven methods fail. Finally, evaluations on hierarchical directed tree networks highlighted the robustness and scalability of the federated architecture, confirming that a single, collaboratively learned shared gain can successfully enforce consensus even in cascaded networks subject to unmodeled heterogeneous parameter perturbations.

\section*{Appendix A: Proof of Lemma \ref{lem:covariance_lipschitz}}

\textit{Proof:} Let the closed-loop system dynamics for the optimal controller and the current controller be denoted as $A_{cl}(K^*_\gamma)$ and $A_{cl}(K_\gamma)$, respectively. The steady-state covariance matrices $X_{K_\gamma^*}^i$ and $X_{K_\gamma}^i$ are the unique positive semi-definite solutions to their respective discrete Lyapunov equations.
Accordingly, the exact perturbation $\Delta X^i_{K_\gamma}=X_{K_\gamma^*}^i - X_{K_\gamma}^i$ is described as, 
\begin{align}
    & \Delta X^i_{K_\gamma} =   A_{cl}(K^*_\gamma) X_{K_\gamma^*}^i A_{cl}(K^*_\gamma)^\top- A_{cl}(K_\gamma) X_{K_\gamma}^i A_{cl}(K_\gamma)^\top = A_{cl}(K_\gamma) \Delta X^i_{K_\gamma} A_{cl}(K_\gamma)^\top + W^i
\end{align}
with $ W^i \triangleq A_{cl}(K^*_\gamma) X_{K_\gamma^*}^i A_{cl}(K^*_\gamma)^\top - A_{cl}(K_\gamma) X_{K_\gamma^*}^i A_{cl}(K_\gamma)^\top.$
By adding and subtracting the cross-term $A_{cl}(K_\gamma) X_{K_\gamma^*}^i A_{cl}(K_\gamma^*)^\top$, $W^i$ factors into,
$$W^i = \left[ A_{cl}(K_\gamma^*) - A_{cl}(K_\gamma) \right] X_{K_\gamma^*}^i A_{cl}(K_\gamma^*)^\top + A_{cl}(K_\gamma) X_{K_\gamma^*}^i \left[ A_{cl}(K_\gamma^*) - A_{cl}(K_\gamma) \right]^\top$$Substituting the matrix difference into this factorization yields:$$W^i = \sqrt{\gamma} \lambda_i B (K_\gamma - K_\gamma^*) X_{K_\gamma^*}^i A_{cl}(K_\gamma^*)^\top + \sqrt{\gamma} \lambda_i A_{cl}(K_\gamma) X_{K_\gamma^*}^i (K_\gamma - K_\gamma^*)^\top B^\top.$$
Applying the triangle inequality and sub-multiplicativity isolates the norms:$$\Vert{}W^i\Vert{}_F \leq \sqrt{\gamma} \lambda_i \Vert{}B\Vert{}_2 \Vert{}K_\gamma - K_\gamma^*\Vert{}_F \Vert{}X_{K_\gamma^*}^i\Vert{}_2 \left( \Vert{}A_{cl}(K_\gamma^*)\Vert{}_2 + \Vert{}A_{cl}(K_\gamma)\Vert{}_2 \right).$$

The spectral norms of the closed-loop matrices are bounded by \eqref{eq:Abound}. Because the optimal cost matrix satisfies $P_{K_\gamma^*}^i \preceq P_{K_\gamma}^i$, the sum of the spectral norms is bounded by $\Vert{}A_{cl}(K_\gamma^*)\Vert{}_2 + \Vert{}A_{cl}(K_\gamma)\Vert{}_2 \leq 2 \sqrt{\frac{\Vert{}P_{K_\gamma}^i\Vert{}_2}{\lambda_{\min}(Q)}}$.
Applying the inequality $\|YZ\|_F \leq \|Y\|_2\|Z\|_F$ yields,
\begin{equation}\label{eq:W_bound_final}
   \Vert{}W^i\Vert{}_F \leq 2 \sqrt{\gamma} \lambda_i \Vert{}B\Vert{}_2 \Vert{}X_{K_\gamma^*}^i\Vert{}_2 \sqrt{\frac{\Vert{}P_{K_\gamma}^i\Vert{}_2}{\lambda_{\min}(Q)}} \Vert{}K_\gamma - K_\gamma^*\Vert{}_F.   
\end{equation}

Next, we evaluate the discrete Lyapunov operator applied to $W^i$.
Because $A_{cl}(K_\gamma)$ is strictly Schur stable, the solution for $\Delta X^i_{K_\gamma}$ can be expressed as an infinite series, which guarantees that the matrix deviation is bounded by,
$$\Vert{}\Delta X^i_{K_\gamma}\Vert{}_F \leq \left( \frac{\Vert{}P_{K_\gamma}^i\Vert{}_2}{\lambda_{\min}(Q)} \right) \Vert{}W^i\Vert{}_F.$$
Finally, substituting the bound for $\|W\|_F$ from \eqref{eq:W_bound_final} establishes the covariance Lipschitz continuity bound,
\begin{equation}
    \|X_{K_\gamma^*}^i - X_{K_\gamma}^i\|_F \leq \frac{2 \sqrt{\gamma}\lambda_{i} \|B\|_2 \|X_{K_\gamma^*}^i\|_2 \|P_{K}^i\|_2^{\frac{3}{2}}}{\lambda_{\min}(Q)^{\frac{3}{2}}} \|K_{\gamma} - K_\gamma^*\|_F.
\end{equation}\hfill $\blacksquare$

\section*{Appendix B: Proof of Lemma \ref{lemma:subopt}}

\textit{Proof:} 
Let $\Delta K = K - K_\gamma^*$. Utilizing the marginal gradient factor $G_K^i = (R + B_i^T P_K^i B_i)K - B_i^T P_K^i A_i$ and $E^i_K=R+B_i^\top P_K^i B_i$, the exact LQR Cost Difference Lemma \cite{fazel2018global} evaluated across the active network modes expands to,
\begin{align*}
  &   J_{\text{con}}(K \mid \gamma) - J_{\text{con},\gamma}^* = \frac{1}{N-1} \sum_{i=2}^N  2 \text{Tr}\left( \Delta K^T G_K^i X_{K_\gamma^*}^i \right) - \frac{1}{N-1} \sum_{i=2}^N \text{Tr}\left( \Delta K^T E_K^i \Delta K X_{K_\gamma^*}^i \right).
\end{align*}
Letting $D = \frac{1}{N-1} \sum_{i=2}^N G_K^i X_{K_\gamma^*}^i$ to represent the aggregated mixed gradient, the cost gap condenses algebraically to,
\begin{align}\label{eq:cost_condensed}
 & J_{\text{con}}(K \mid \gamma) - J_{\text{con},\gamma}^* =   2 \text{Tr}\left( \Delta K^T D \right) - \frac{1}{N-1} \sum_{i=2}^N \text{Tr}\left( \Delta K^T E_K^i \Delta K X_{K_\gamma^*}^i \right).
\end{align}
By the properties of the trace operator,
\begin{align}
  \nonumber  & \text{Tr}\left( \Delta K^T E_K^i \Delta K X_{K_\gamma^*}^i \right) \geq \lambda_{\min}(X_{K_\gamma^*}^i) \text{Tr}\left( \Delta K^T E_K^i \Delta K \right) \geq \lambda_{\min}(R)\lambda_{\min}(X_{K_\gamma^*}^i) \|\Delta K\|_F^2.
\end{align}
Averaging this lower bound across all active modes and substituting it back into \eqref{eq:cost_condensed} yields a bounded quadratic form,
\begin{align}
    &  J_{\text{con}}(K \mid \gamma) - J_{\text{con},\gamma}^* \leq  2 \text{Tr}\left( \Delta K^T D \right) - \lambda_{\min}(R)\min_i \lambda_{\min}(X_{K_\gamma^*}^i)  \|\Delta K\|_F^2.
\end{align}
   
To analytically eliminate the gain deviation $\Delta K$, let $\alpha = \lambda_{\min}(R)\min_i \lambda_{\min}(X_{K_\gamma^*}^i)$. The algebraic expansion is of the form $2 \text{Tr}\left( \Delta K^T D \right) - \alpha \|\Delta K\|_F^2$ which takes its maximum at $\Delta K = \frac{1}{\alpha} D$. Accordingly,
\begin{align} \label{eq:grad_dom_base}
   \nonumber &J_{\text{con}}(K \mid \gamma) - J_{\text{con},\gamma}^* \leq \frac{1}{\alpha} \|D\|_F^2  = \frac{1}{\lambda_{\min}(R)\min_i \lambda_{\min}(X_{K_\gamma^*}^i)} \left\| \frac{1}{N-1} \sum_{i=2}^N G_K^i X_{K_\gamma^*}^i \right\|_F^2.
\end{align}
Next, 
\begin{align*}
    & \frac{1}{N-1} \sum_{i=2}^N G_K^i X_{K_\gamma^*}^i = \left( \frac{1}{N-1} \sum_{i=2}^N G_K^i X_K^i \right)  + \left( \frac{1}{N-1} \sum_{i=2}^N G_K^i (X_{K_\gamma^*}^i - X_K^i) \right).
\end{align*}
The first term is precisely the true policy gradient of the aggregate discounted consensus cost, $\nabla J_{\text{con}}(K \mid \gamma)$, and the second term is the covariance mismatch residual, $\mathcal{X}(K, K_\gamma^*)$. Substituting these identities into our norm bound yields,
\begin{align*}
    &   J_{\text{con}}(K \mid \gamma) - J_{\text{con},\gamma}^* \leq  \frac{1}{\lambda_{\min}(R)\min_i \lambda_{\min}(X_{K_\gamma^*}^i)} \left\| \nabla J_{\text{con}}(K \mid \gamma) + \mathcal{X}(K, K_\gamma^*) \right\|_F^2.
\end{align*}
Finally, applying the algebraic norm inequality $\Vert{}A + B\Vert{}_F^2 \leq (1 + \epsilon)\Vert{}A\Vert{}_F^2 + \left(1 + \frac{1}{\epsilon}\right)\Vert{}B\Vert{}_F^2$ for all $\epsilon > 0$ we distribute the explicit conditioning constants to obtain the final network performance bound,
\begin{align*}
    & J_{\text{con}}(K \mid \gamma) - J_{\text{con},\gamma}^* \leq  \frac{1+\epsilon}{\lambda_{\min}(R)\min_i \lambda_{\min}(X_{K_\gamma^*}^i)} \left\| \nabla J_{\text{con}}(K \mid \gamma) \right\|_F^2 + \frac{1+\epsilon^{-1}}{\lambda_{\min}(R)\min_i \lambda_{\min}(X_{K_\gamma^*}^i)} \left\| \mathcal{X}(K, K_\gamma^*) \right\|_F^2.
\end{align*}
This completes the proof. \hfill $\blacksquare$

\bibliographystyle{unsrt}
\bibliography{MyReferences}  

@article{feng2021consensusability,
  title={Consensusability and global optimality of discrete-time linear multiagent systems},
  author={Feng, Tao and Zhang, Jilie and Tong, Yin and Zhang, Huaguang},
  journal={IEEE Transactions on Cybernetics},
  volume={52},
  number={8},
  pages={8227--8238},
  year={2021},
  publisher={IEEE}
}

@article{farjadnasab2022model,
  title={Model-free {LQR} design by {Q}-function learning},
  author={Farjadnasab, Milad and Babazadeh, Maryam},
  journal={Automatica},
  volume={137},
  pages={110060},
  year={2022},
  publisher={Elsevier}
}

@book{west.2001,
  title={Introduction to graph theory},
  author={West, Douglas Brent and others},
  volume={2},
  year={2001},
  publisher={Prentice hall Upper Saddle River}
}

@article{feng2021q,
  title={{Q}-learning algorithm in solving consensusability problem of discrete-time multi-agent systems},
  author={Feng, Tao and Zhang, Jilie and Tong, Yin and Zhang, Huaguang},
  journal={Automatica},
  volume={128},
  pages={109576},
  year={2021},
  publisher={Elsevier}
}

@article{dutta2022strict,
  title={Strict {L}yapunov functions for dynamic consensus in linear systems interconnected over directed graphs},
  author={Dutta, Maitreyee and Panteley, Elena and Loria, Antonio and Sukumar, Srikant},
  journal={IEEE Control Systems Letters},
  volume={6},
  pages={2323--2328},
  year={2022},
  publisher={IEEE}
}

@article{yang2020cooperative,
  title={Cooperative output tracking of unknown heterogeneous linear systems by distributed event-triggered adaptive control},
  author={Yang, Ruohan and Liu, Lu and Feng, Gang},
  journal={IEEE Transactions on Cybernetics},
  volume={52},
  number={1},
  pages={3--15},
  year={2022},
  publisher={IEEE}
}

@article{cao2023distributed,
  title={Distributed adaptive output consensus of unknown heterogeneous non-minimum phase multi-agent systems},
  author={Cao, Wenji and Liu, Lu and Feng, Gang},
  journal={IEEE/CAA Journal of Automatica Sinica},
  volume={10},
  number={4},
  pages={997--1008},
  year={2023},
  publisher={IEEE}
}

@article{zhang2023data,
  title={Data-driven control of consensus tracking for discrete-time multi-agent systems},
  author={Zhang, Xiufeng and Wang, Gang and Sun, Jian},
  journal={Journal of the Franklin Institute},
  volume={360},
  number={7},
  pages={4661--4674},
  year={2023},
  publisher={Elsevier}
}

@article{de2019formulas,
  title={Formulas for data-driven control: Stabilization, optimality, and robustness},
  author={De Persis, Claudio and Tesi, Pietro},
  journal={IEEE Transactions on Automatic Control},
  volume={65},
  number={3},
  pages={909--924},
  year={2019},
  publisher={IEEE}
}

@book{lancaster1995algebraic,
  title={Algebraic {R}iccati equations},
  author={Lancaster, Rodman and Lancaster Peter},
  year={1995},
  publisher={Oxford Oxford University Press}
}

@article{perdomo2021stabilizing,
  title={Stabilizing dynamical systems via policy gradient methods},
  author={Perdomo, Juan and Umenberger, Jack and Simchowitz, Max},
  journal={Advances in neural information processing systems},
  volume={34},
  pages={29274--29286},
  year={2021}
}

@inproceedings{fazel2018global,
  title={Global convergence of policy gradient methods for the linear quadratic regulator},
  author={Fazel, Maryam and Ge, Rong and Kakade, Sham and Mesbahi, Mehran},
  booktitle={International conference on machine learning},
  pages={1467--1476},
  year={2018},
  organization={PMLR}
}

@inproceedings{malik2019derivative,
  title={Derivative-free methods for policy optimization: Guarantees for linear quadratic systems},
  author={Malik, Dhruv and Pananjady, Ashwin and Bhatia, Kush and Khamaru, Koulik and Bartlett, Peter and Wainwright, Martin},
  booktitle={The 22nd international conference on artificial intelligence and statistics},
  pages={2916--2925},
  year={2019},
  organization={PMLR}
}

@article{TCNS2026,
  title={Model-Free Dynamic Consensus in Multi-Agent Systems: A Q-Function Perspective},
  author={Babazadeh, Maryam and Bajcinca, Naim},
  journal={IEEE Transactions on Control of Network Systems},
  year={2026},
  publisher={IEEE}
}

@inproceedings{fujinami2025policy,
  title={Policy gradient for {LQR} with domain randomization},
  author={Fujinami, Tesshu and Lee, Bruce D and Matni, Nikolai and Pappas, George J},
  booktitle={2025 IEEE 64th Conference on Decision and Control (CDC)},
  pages={4174--4181},
  year={2025},
  organization={IEEE}
}

@article{wang2026model,
  title={Model-free learning with heterogeneous dynamical systems: A federated {LQR} approach},
  author={Wang, Han and Toso, Leonardo F and Mitra, Aritra and Anderson, James},
  journal={Transactions on Machine Learning Research},
  year={2026}
}

@article{ren2020federated,
  author    = {Ren, Z. and Zhong, A. and Zhou, Z. and Li, N.},
  title     = {Federated {LQR}: Learning through sharing},
  journal   = {arXiv preprint arXiv:2011.01815},
  year      = {2020},
  doi       = {10.48550/arXiv.2011.01815}
}

@article{hu2023toward,
  title={Toward a theoretical foundation of policy optimization for learning control policies},
  author={Hu, Bin and Zhang, Kaiqing and Li, Na and Mesbahi, Mehran and Fazel, Maryam and Ba{\c{s}}ar, Tamer},
  journal={Annual Review of Control, Robotics, and Autonomous Systems},
  volume={6},
  number={1},
  pages={123--158},
  year={2023},
  publisher={Annual Reviews}
}

@article{mohammadi2021convergence,
  title={Convergence and sample complexity of gradient methods for the model-free linear--quadratic regulator problem},
  author={Mohammadi, Hesameddin and Zare, Armin and Soltanolkotabi, Mahdi and Jovanovi{\'c}, Mihailo R},
  journal={IEEE Transactions on Automatic Control},
  volume={67},
  number={5},
  pages={2435--2450},
  year={2021},
  publisher={IEEE}
}

@inproceedings{lamperski2020computing,
  title={Computing stabilizing linear controllers via policy iteration},
  author={Lamperski, Andrew},
  booktitle={2020 59th IEEE Conference on Decision and Control (CDC)},
  pages={1902--1907},
  year={2020},
  organization={IEEE}
}

@article{lu2024federated,
  title={Federated learning with non-iid data: A survey},
  author={Lu, Zili and Pan, Heng and Dai, Yueyue and Si, Xueming and Zhang, Yan},
  journal={IEEE Internet of Things Journal},
  volume={11},
  number={11},
  year={2024}
}

@article{baduwal2026federated,
  title={Federated learning: A survey of core challenges, current methods, and opportunities},
  author={Baduwal, Madan and Paudel, Priyanka and Chaudhary, Vini},
  journal={Computers},
  volume={15},
  number={3},
  year={2026}
}

@article{woo2025blessing,
  title={The blessing of heterogeneity in federated Q-learning: Linear speedup and beyond},
  author={Woo, Jiin and Joshi, Gauri and Chi, Yuejie},
  journal={Journal of Machine Learning Research},
  volume={26},
  number={26},
  year={2025}
}

@article{rostami2026scalar,
  title={Scalar Federated Learning for Linear Quadratic Regulator},
  author={Rostami, Mohammadreza and Talebi, Shahriar and Kia, Solmaz S},
  journal={arXiv preprint},
  year={2026}
}

@inproceedings{toso2024asynchronous,
  author    = {Toso, L. F. and Wang, H. and Anderson, J.},
  title     = {Asynchronous heterogeneous linear quadratic regulator design},
  booktitle = {2024 IEEE 63rd Conference on Decision and Control},
  year      = {2024}
}






\end{document}